\documentclass[journal, 10pt,final]{IEEEtran}

\makeatletter

\usepackage{comment}
\let\wfs@comment@comment\comment
\let\comment\@undefined

\usepackage[final]{changes}
\let\wfs@changes@comment\comment
\let\comment\@undefined

\newcommand\comment{%
    \ifthenelse{\equal{\@currenvir}{comment}}
    {\wfs@comment@comment}
    {\wfs@changes@comment}%
}

\makeatother

\definechangesauthor[name={Jose Terrero}, color=red]{JT}
\definechangesauthor[name={Yuqi Fu}, color=green]{yuqi}

\usepackage{algorithm,multirow}
\usepackage[noend]{algorithmic}
\usepackage{graphicx}
\usepackage{cite}
\usepackage{multirow}
\usepackage{amssymb,amsthm,mathrsfs}
\usepackage{amsfonts,epsf,graphicx,times,amsmath,multirow,url,cite}
\usepackage{float}
\usepackage{flexisym}
\usepackage[capitalise]{cleveref}
\usepackage{float}
\usepackage{verbatim,epstopdf}

\usepackage{amsthm}
\usepackage{subcaption}
\usepackage{color}

\usepackage{footnote}
\usepackage{footmisc}
\makesavenoteenv{tabular}
\makesavenoteenv{table}
\renewcommand{\thefootnote}{\alph{footnote}}
\newcommand{\astfootnote}[1]{
\let\oldthefootnote=\thefootnote
\setcounter{footnote}{0}
\renewcommand{\thefootnote}{\fnsymbol{footnote}}
\footnote{#1}
\let\thefootnote=\oldthefootnote
}

\usepackage{wrapfig}

\graphicspath{{figs/}}
	\DeclareGraphicsExtensions{.eps}

\usepackage{array}
\newcolumntype{L}[1]{>{\raggedright\let\newline\\\arraybackslash\hspace{0pt}}m{#1}}
\newcolumntype{C}[1]{>{\centering\let\newline\\\arraybackslash\hspace{0pt}}m{#1}}
\newcolumntype{R}[1]{>{\raggedleft\let\newline\\\arraybackslash\hspace{0pt}}m{#1}}

\makeatletter
\def\ps@headings{%
\def\@oddhead{\mbox{}\scriptsize\rightmark \hfil \thepage}%
\def\@evenhead{\scriptsize\thepage \hfil \leftmark\mbox{}}%
\def\@oddfoot{}%
\def\@evenfoot{}}
\makeatother
\usepackage [english]{babel}
\usepackage [autostyle, english = american]{csquotes}
\MakeOuterQuote{"}

\usepackage{caption}
\usepackage{subcaption}

\begin{document}

\date{}


\title{\huge Resource Management Schemes for Cloud Native Platforms with Computing Containers of Docker and Kubernetes}
\author{

Ying Mao,~\IEEEmembership{Member,~IEEE,}  Yuqi Fu Suwen Gu,~\IEEEmembership{Student Member,~IEEE,} Wenrui Mu
\\ Long Cheng, Qingzhi Liu,~\IEEEmembership{Member,~IEEE}

\IEEEcompsocitemizethanks{

\IEEEcompsocthanksitem Y. Mao, Y. Fu, S. Gu, and W. Mu are with the Department of Computer and Information Science at Fordham University in the New York City. E-mail: \{ymao41, yfu81, sgu21, svhaduri\}@fordham.edu
\IEEEcompsocthanksitem L. Cheng is with the School of Computing, Dublin City University, Ireland. E-mail: long.cheng@dcu.ie
\IEEEcompsocthanksitem Q. Liu is with Information Technology Group, Wageningen Univer- sity, The Netherlands E-mail: qingzhi.liu@wur.nl

}
}

\maketitle

\begin{abstract}

Businesses have made increasing adoption and incorporation of cloud technology into internal processes in the last decade. The cloud-based deployment provides on-demand availability without active management. More recently, the concept of cloud-native application has been proposed and represents an invaluable step toward helping organizations develop software faster and update it more frequently to achieve dramatic business outcomes. Cloud-native is an approach to build and run applications that exploit the cloud computing delivery model's advantages. It is more about how applications are created and deployed than where. The container-based virtualization technology, e.g., Docker and Kubernetes, serves as the foundation for cloud-native applications. This paper investigates the performance of two popular computational-intensive applications, big data and deep learning, in a cloud-native environment. We analyze the system overhead and resource usage for these applications. Through extensive experiments, we show that the completion time reduces by up to 79.4\% by changing the default setting and increases by up to 96.7\% due to different resource management schemes on two platforms. Additionally, the resource release is delayed by up to 116.7\% across different systems. Our work can guide developers, administrators, and researchers to better design and deploy their applications by selecting and configuring a hosting platform.

\end{abstract}

\begin{IEEEkeywords}
 Container Management; Docker Swarm; Kubernetes; Tensorflow; Pytorch; Apache Yarn and Spark;
\end{IEEEkeywords}


\section{Introduction}
\label{intro}

In the past decade, businesses \replaced[id=JT]{from all over the world, regardless of size or industry, have been turning to cloud services to fulfill their computing needs.}{ of all sizes, industries, and geographies are turning to cloud services.} 
\added[id=JT]{Among the benefits of embracing cloud services at a corporate level are that}
moving to cloud computing can reduce the cost of managing and maintaining the IT systems \deleted[id=JT]{for many of them.}
\added[id=JT]{and those} \deleted[id=JT]{The} cloud infrastructures can be scaled up or down based on the computation and storage needs quickly to suit the business development, allowing flexibility as the needs change.  

According to a Forbes report~\cite{forbes}, spending on cloud computing infrastructure and platforms \replaced[id=JT]{grew}{ will grow} at a 30\% compound annual growth rate from 2013 through 2018 compared with 5\% percent growth for overall enterprise IT. 
From the cloud computing services providers \replaced[id=JT]{perspective}{ prospective}, such as Amazon Web Service~\cite{aws}, Microsoft Azure~\cite{azure} and Google Cloud Platform~\cite{gcp}, \deleted[id=JT]{a} virtual machine backed virtualization technology powers the datacenters and provides services to multiple users on the same server while maintains isolation among them.
From the \replaced[id=JT]{client's}{ clients'} \replaced[id=JT]{perspective,}{ prospective,} firstly, they develop and test applications locally and then, create a virtual machine on the cloud, which is identical to their run-time environment, e.g., operating system and hardware settings, and finally, deploy the applications inside the virtual machines to provide the services.

In a small-scale system, the system is efficient since clients can purchase to replicate the active virtual machines when their demands are increasing and \replaced[id=JT]{terminate}{ close} the virtual machines when they need to scale down. 
As the system grows, however, providing services through virtual machines \replaced[id=JT]{implies}{ means} running many duplicated instances of the same OS and redundant boot volumes, which lead to large scheduling overhead. Additionally, virtual machines usually target providing full services, e.g. access control, database, and logging on one type of machine. Although replicating virtual machines can improve the performance of database services when the workload of database queries is increasing, the whole system will inevitably waste \replaced[id=JT]{some}{ part} of the resources that are assigned for other services on the same virtual machine.

\begin{wrapfigure}{l}{0.5\linewidth}
\centering
\includegraphics[width=\linewidth]{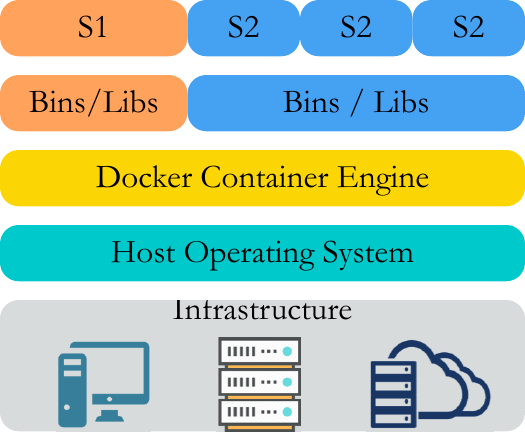}
\caption{Docker Containers}
\label{fig:docker-standalone} 
\end{wrapfigure}
To address the above limitations and fully utilize the advantages of cloud computing, \replaced[id=JT]{containerization tools}{ containers}, such as Docker~\cite{docker}, are designed for deploying and running distributed applications without launching entire virtual machines. Instead, multiple isolated containers \replaced[id=JT]{share}{ are sharing} the host operating system and physical resources. Furthermore, \replaced[id=JT]{containerization}{ containers} \replaced[id=JT]{is}{ are} \replaced[id=JT]{service}{ services} level virtualization \added[id=JT]{in} that each \replaced[id=JT]{container}{ of them} focuses on providing a limited number of services, which makes the system more flexible for the elasticity of demands.

Fig.~\ref{fig:docker-standalone} shows a machine \added[id=JT]{that} hosts four containers, which target on providing two services (S1 and S2). \deleted[id=JT]{As we can see,} Instead of loading the whole guest operating system, \replaced[id=JT]{Docker}{ it} only installs the required binaries and libraries, which saves \replaced[id=JT]{significant amounts}{ lots} of resources and \replaced[id=JT]{greatly speeds up}{speedup} the booting process. Currently, S1 is offered by one container, while S2 \replaced[id=JT]{is offered by}{ has} two. If S1's workload \replaced[id=JT]{increases}{ is increasing}, the system can initiate a new container for S1 or, depending on S2's situation, it may stop one of S2's containers and \replaced[id=JT]{direct}{ move} the resource \added[id=JT]{towards} to S1.

With \deleted[id=JT]{the} container-based virtualization, the \replaced[id=JT]{concept}{concepts} of cloud-native \replaced[id=JT]{is}{ are} proposed to develop applications built \replaced[id=JT]{as}{ with} services \added[id=JT]{that are} packaged in virtualized containers, deployed as microservices and managed on elastic infrastructure through agile development, \deleted[id=JT]{and} operation processes and continuous delivery workflows.
In a production environment of cloud-native applications, a multi-node cluster is used to provide the backend infrastructures for running at scale. In the domain of managing clusters \replaced[id=JT]{for containerization}{ of containers}, Docker Swarm~\cite{dockerswarm} and Kubernetes~\cite{k8s} are the key players. Both of them provide rich functionalities for resource management schemes, security \replaced[id=JT]{policy enforcement}{ policies}, network access, system \replaced[id=JT]{administration}{ monitor}, etc.
However, these  functionalities would introduce overhead to the system. In this paper, we investigate the performance of using cloud-native frameworks from the perspective of resource management. 
The main contributions of this paper are summarized as follows,

\begin{itemize}
\item We build a monitor system to collect real-time resource usage data as well as calculating completion time on Docker and Kubernetes platforms for containerized big data and deep learning applications.
\item We conduct intensive experiments on each individual platform with different workloads and discover that the completion time can be reduced for up to 79.4\% and 69.4\% by changing the default configurations on Docker and Kubernetes, respectively.
\item Through extensive experiments, we demonstrate that the Docker platform delayed the resource release process for up to 116.7\% for short-lived deep learning jobs. Furthermore, the completion time can increase up to 96.7\% when compared with each other due to different resource management schemes.
\end{itemize}

 The remainder of this paper is organized as follows. In Section~\ref{rel}, we review the related works in the literature. In Section~\ref{back}, we introduce the cloud-native applications as well as two representative hosting platforms, Docker and Kubernetes, along with their resource management schemes. Section~\ref{eval} presents the comprehensive experiments that we conducted to evaluate the platforms and  the analytical results that generated by the applications. Finally, Section~\ref{con} concludes the paper.

\section{Related Work}
\label{rel}
Cloud computing has been commercialized and started serving the industry over a decade~\cite{awshistory}.
Whenever we scroll through Facebook~\cite{fb} or Twitter~\cite{twitter}, watch streaming videos, access an online email service like Gmail~\cite{gmail} or Outlook~\cite{outlook}, or use the applications on our phones, we are accessing data and services on the cloud. 
Based on cloud infrastructures, many research projects have been conducted to optimize system performance~\cite{chen2020woa, mao2017draps,mao2014pasa,mao2018dress} and enable new applications~\cite{acharya2019workload,acharya2019edge, harvey2017edos}. 
Cloud computing is, traditionally, powered by virtual machines in order to share the resources among different clients and maintain an isolated environment for each of them. While this architecture works great for a small scale system, it suffers from large overhead on service provision, storage spaces, and system elasticity~\cite{xu2013managing} when serving a larger and more complicated system.

More recently, the concept of cloud-native computing is proposed for architecting applications specifically to run in the elastic and distributed nature, which required by modern cloud computing platforms. At the back end side, container-based virtualization technologies serve as the foundation of cloud-native applications. Comparing with virtual machines, containers are more lightweight, resource-friendly, and efficient to scale up~\cite{felter2015updated}. 
There are lots of works investigate the performance of containers and virtual machines for various perspectives. 
A comparative study of different virtualization technologies is conducted in~\cite{sharma2016containers}, the authors 
conclude that while containers may offer bare-metal performance, they are lack of rich isolation properties.
ContainerCloudSim~\cite{piraghaj2017containercloudsim} extends CloudSim for modeling and simulation of containerized cloud computing environments.  
The authors in~\cite{tay2017performance} present a mathematical framework for decision making around placing and migrating workloads in a data-center where applications are packaged as OS containers running on virtual machines. 
In the domain of big data processing, experiments in~\cite{bhimani2017accelerating, zhang2018comparative} demonstrate that containers enhance the performance of enterprise clouds. 
  
Besides benefits offered by individual containers, a more advanced part of cloud-native computing is container orchestration. In this domain, Docker Swarm~\cite{dockerswarm} and Kubernetes~\cite{k8s} are the key players. 
Asif et. al provides a non-exhaustive and prescriptive guide in~\cite{khan2017key} to identifying and implementing key mechanisms required in a container orchestration platform. Eddy et. al study the e performance overhead of container orchestration frameworks for running and managing database clusters. NBWGuard~\cite{xu2018nbwguard} is proposed as a design for network bandwidth management for Kubernetes.
With emerging containerized deep learning applications, authors in~\cite{fu2019progress} propose ProCon, a start-of-art container scheduler that takes the progress of each job into resource allocation. Furthermore, in order to optimize the container management platform for the model training process, TRADL~\cite{zheng2019target} and FlowCon~\cite{zheng2019flowcon} provide efficient scheduling algorithms to boost the scalability of the systems. 

In addition to the existing research, in this paper, we investigate the resource usage patterns of cloud-native applications in two popular types, deep learning framework (e.g. Tensorflow~\cite{tensorflow} and Pytorch~\cite{pytorch}) and big data processing~\cite{spark}.


 \section{Applications on the cloud}
 \label{back}
In this section, we will discuss cloud-native applications and container technologies, such as Docker and Kubernetes, which are the fundamental building block of cloud-native applications. 

\subsection{Cloud Native}
Cloud-native is generally an approach to build and run applications that leverage the benefits of the cloud computing model. The Cloud Native Computing Foundation (CNCF)~\cite{cncf} characterizes cloud-native as using open source software stack with the following properties.  
\begin{itemize}
\item Containerized, which means each part of the application is packaged into its own container. A property that enables reproducibility, transparency, and resource isolation.
\item Dynamically orchestrated. Containers are actively scheduled and regulated to optimize resource utilization. 
\item Microservices-oriented. Applications are divided into microservices and managed on elastic infrastructure to improve the overall agility and maintainability of applications. 
\end{itemize}


A container is a standard unit of software that packages up the programs and all their dependencies together. Containers isolate an application from its environment and guarantee that it works reliably across different environments. Cloud-native applications, on the other hand, are a collection of independent services that are packaged as lightweight containers.



\begin{figure}[ht]
\centering
\includegraphics[width=0.55\linewidth]{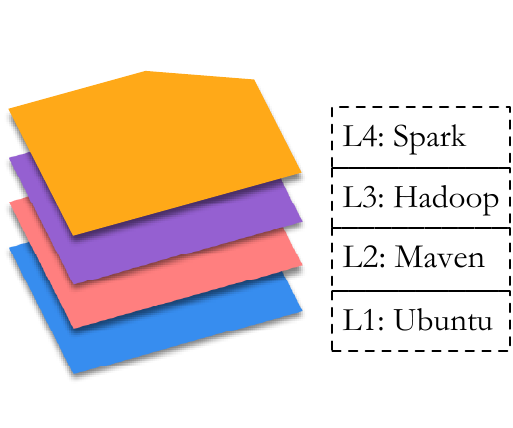}
\caption{Sample Container Images}
\label{fig:container-images} 
\end{figure}

Different from virtual machines, containers virtualize operating systems instead of hardware, which makes them more portable, efficient, and scalable. Containers are an abstraction at the application layers and multiple containers can run on the same machine while each running as an isolated process. 
The images of containers have a layered structure. Fig.~\ref{fig:container-images} presents a sample image for running 
a container of Apache Spark. We first pull the Ubuntu Operating System (L1) as the base from DockerHub~\cite{dockerhub}.
On top of the OS layer, we load the maven (L2) for Java-related packages and then, Hadoop HDFS (L3) is built on top of L2.
Finally, we fetch Spark (L4) for data processing.
Compare to virtual machines, containers occupy less space, can handle more applications with fewer hardware resources. 

\subsection{Container Orchestration}
Cloud-native applications are developed as a system of multiple, relatively small microservices that work together to provide a comprehensive system. Services could be developed and deployed independently where each realizes exactly one functionality so when change is required in a certain part of the application, only the related services need to be modified and redeployed. Besides, the services of a cloud-native application could use different types of languages, runtimes, and frameworks. Each service is \replaced{could be} developed using the language and framework that best suited for specific functionality. 

Manual application deployment and management could be challenging, because monitoring the system status, managing active containers, and balancing the load between running applications at the same time is costly and respond slowly to unexpected changes in the system. Cloud-native applications use automatically container orchestration tools such as Docker Swarm and Kubernetes to be fully benefited from a cloud platform. 

\subsubsection{Docker Swarm Mode}

While Docker creates virtualized containers on single machines, Docker Swarm Mode is an open-source container orchestration platform, which provides a suite of management toolkit for a cluster of containers. It is the native
clustering solution for Docker containers, which has the advantage of being firmly integrated
into the Docker ecosystem and utilizing its own API. A swarm consists of multiple Docker
hosts that run in swarm mode. 

\begin{figure}[ht]
\centering
\includegraphics[width=1\linewidth]{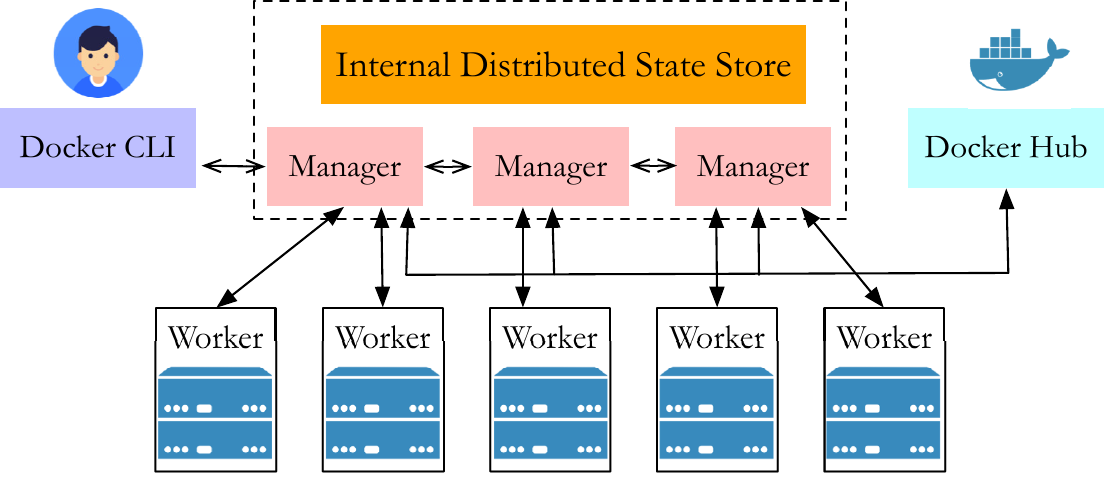}
\caption{Docker Swarm Mode}
\label{fig:dockerswarm} 
\end{figure}

Fig.~\ref{fig:dockerswarm} illustrates the system architecture of Docker swarm mode 
in a multi-node cluster. 
There are two types of nodes in this cluster, worker nodes, and manager nodes. Worker nodes are responsible for running tasks; on the other hand, manager nodes accept specifications from the user and pull images from sources (e.g. Docker Hub).
Worker nodes receive and execute tasks from
managers. The agent on the worker node
reports the state of tasks to the manager, who is
responsible for reconciling the desired state with the actual cluster state. 
Services define how individual containers
can distribute themselves across the nodes. 
The Task is a container and the commands to be run inside of
the container. Each manager is in charge of one or multiple workers. The number of managers 
can increase along with the workload.

\subsubsection{kubernetes}

Kubernetes is an open-source platform created by Google for container deployment
operations, scaling, and management of a containerized application. It is a management tool
that has been specifically designed to simplify the scalability of workload using containers.
Kubernetes supports the same images of Docker.

\begin{figure}[ht]
\centering
\includegraphics[width=1\linewidth]{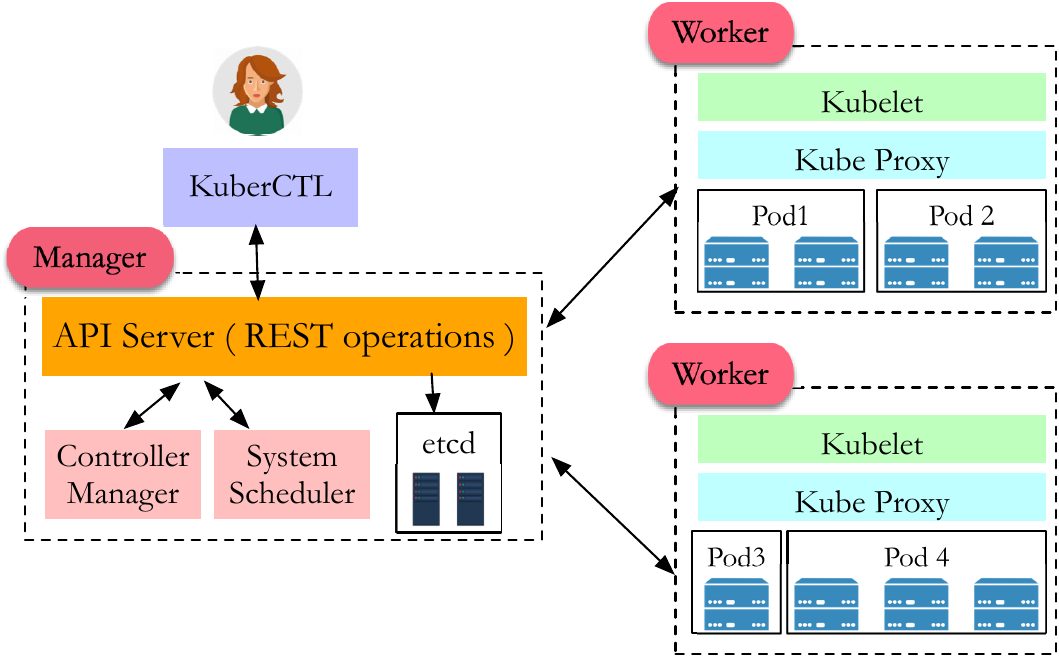}
\caption{Kubernetes}
\label{fig:k8s} 
\end{figure}

Fig~\ref{fig:k8s} plots the architecture of a Kubernetes cluster. Similar to Docker Swarm, 
the system has managers and workers. The clients can use kubectl, a command-line interface for running commands against Kubernetes clusters, to interact with API server, which resides in the manager.
The manager node is responsible for the Kubernetes
cluster and takes care of orchestrating
the worker nodes by using the controller and scheduler. It utilizes etcd for the services of 
consistent and highly-available key-value stores.
On the worker side, kebulet is running on each node in the cluster. It maintains the status of containers on this worker.
A group of containers, one or multiple, on the same node that is created, scheduled, and deployed together
is called a pod. This group of containers would share storage,
namespace, cgroups, IP addresses. 
Kubernetes also introduces the concept of a service,
which is an abstraction on top of a number of pods, typically requiring to run a proxy, for
other services to communicate with it via virtual IP addresses. 
Once the system is online, workers communicate with managers for configurations of containers
downloading the images and starting the containers.

\subsection{Resource Management for Cloud Native Applications}

In a busy cloud computing platform, the resource contention happens all the time. 
From the operating system's point of view, the active containers are not different from regular processes, where
the resource allocation is controlled by the system. Considering a cloud-native application, which powered by
a cluster of containers, two levels of resource management schemes exist: individual worker level and container placement level.

\subsubsection{Individual Workers}
Different from fixed resource configurations for virtual machines, containers employ various plans for resource management.

\noindent {\bf Docker Platform}: When initiating, no resource constraints, by default, are applied to newly created containers. 
In this scenario, each one can use as much of a given resource as the host's kernel scheduler allows. Multiple containers
compete with resources freely in the system.
When necessary, developers have the option to start one with specific resource limits. For example, the command, \texttt{docker run -it --cpus=".7" spark /bin/bash}, guarantees the new container at most 70\% of the CPU (one core) every second. 
Please note that Docker utilizes the "soft limit" concept, where the value serves as the upper limit to containers and if the actual usage can not top-up the limits, the reminding resources could be used by others.
In addition to preset constraints, users can update the configuration at runtime. For instance, the command, \texttt{docker run -dit --name spark --memory 2048M}, dynamically limit the memory usage of a "spark" container to 2G.
The supported resource types are CPU, memory, disk I/O, and network I/O.

\noindent {\bf Kubernetes Platform}: Similar to Docker, when clients specify a Pod, they can optionally specify how much CPU and memory each container needs, in terms of limits and requests.
A limit is the maximum amount of resources that Kubernetes will allow the container to use.
\begin{wrapfigure}{l}{0.4\linewidth}
\centering
\includegraphics[width=\linewidth]{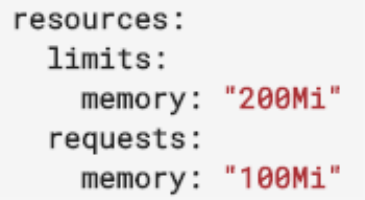}
\caption{YAML}
\label{fig:yaml} 
\end{wrapfigure}
A request is the amount of that resources that the system will guarantee for the container, and Kubernetes will use this value to decide on which node to place the pod.
For example, Fig.~\ref{fig:yaml} shows a piece of YAML~\cite{yaml} file that the developer specified for memory spaces.

Kubernetes employs three resource assignment policies, which aim to provide different Quality of Service (QoS) classes.
\begin{itemize}
\item Guaranteed: Every Container in the Pod must have memory and CPU limit as well as a memory and CPU request, and they must be the same.
\item Best Effort: If a Pod does not have any memory or CPU limits or requests, it is in the "best-effort" mode, where it competes for resources freely with others.
\item Burstable: The Pod does not meet the criteria for Guaranteed and at least one container in the Pod has a memory or CPU request.
\end{itemize}

\subsubsection{Container placement}

In a cluster, typically, there are multiple workers serve as computing nodes. 
For an incoming job request, the manager has to decide which worker should host the task.
Because workers' status, such as the number of active jobs and resource availabilities varies, 
different container placement strategies lead to various performance.

\noindent {\bf Docker Swarm}: There are three container placement schemes. (1) Spread (default) aims to place a container
on the node with the fewest running containers; (2) Binpack places a container onto the most packed node in the
cluster; (3) Random, randomly pick up a worker as the host.
 
\noindent {\bf Kubernetes}: Usually,  Kubernetes scheduler automatically determines an appropriate node for the pod with a scoring algorithm, which calculates a score to every worker based on multiple factors such as available resources. 
However, there are situations when more control is needed for the placement. A pod can request a node with specific characteristics through node affinity and it is also possible to ensure that a set of pods are placed on the same node to avoid latency. 
Placement strategies can be implemented by a set of labels and annotations appropriately included or excluded within pods and nodes.


\section{Performance Evaluation}
\label{eval}

In this section, we explore two popular and representative container platforms, Docker Swarm and Kubernetes. 

\begin{figure}[ht]
\centering
\includegraphics[width=0.90\linewidth]{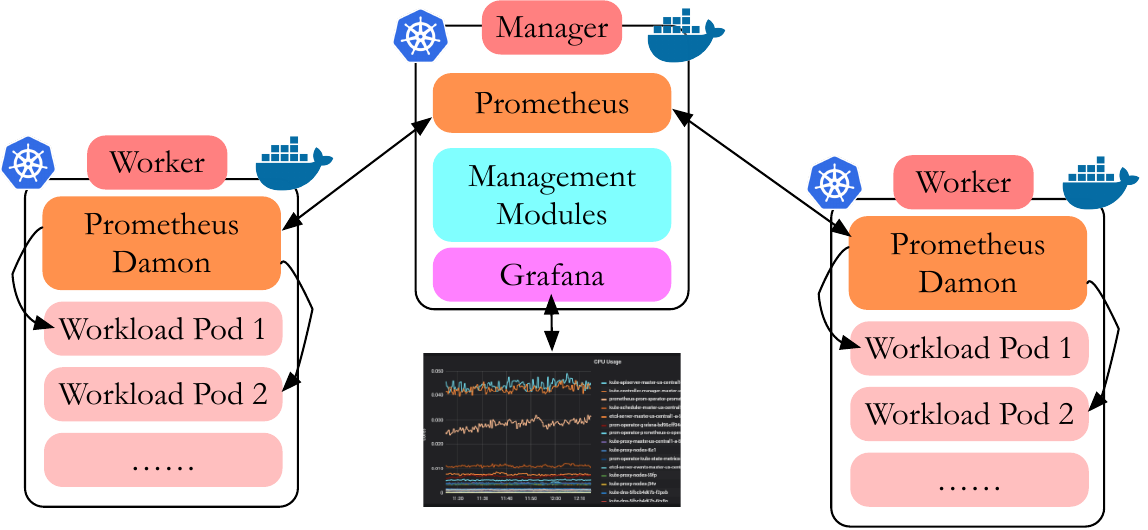}
\caption{Resource Monitoring}
\label{fig:monitor} 
\end{figure}

\subsection{Experiment Setup, Workloads and Evaluation Metrics}

\subsubsection{Monitoring the resources} 
We build a container monitor system based on Prometheus~\cite{prometheus} and Grafana~\cite{grafana}.
Fig.~\ref{fig:monitor} shows the system that we used for our experiments. 
In this system, Prometheus daemon is running on each worker to collect the resource usage data of all the containers/pods on this node. The data is collected by the managers and analyzed, presented by Grafana.

\subsubsection{Workloads and Testbeds}

In the experiments, we mainly evaluate two types of applications, big data (BD) processing jobs, and deep learning (DL) training jobs.
For big data processing, we evaluate Apache Yarn~\cite{yarn} and Apache Spark~\cite{spark}, which are popular distributed general-purpose big data processing framework. 
For the deep learning applications, we focus on two frameworks, Tensorflow~\cite{tensorflow} and Pytorch~\cite{pytorch} that are widely adopted for training deep learning models in the industry. 

Table~\ref{table:workload} presents the experiments that we conduct for evaluating the container orchestration toolkits.

\begin{table}[ht]\small
	\centering
	\caption{Workload for Experiments}
	\scalebox{0.95}{
		
		\begin{tabular}{ | c | c | c | }				
			\hline
			Application    & Platform&  Category  \\ \hline	
			Variational Autoencoders &	Tensorflow	 & DL  \\ \hline	
			MNIST - CNN   &  Tensorflow      & DL  \\ \hline
			MNIST - RNN & Tensorflow & DL \\ \hline
			MNIST - Identity MRNN & Pytorch & DL \\ \hline
			VAE - MLP & Pytorch & DL \\ \hline\hline
			
			PageRank & Spark & BD \\ \hline
			WordCount &  Yarn & BD \\ \hline 
						
		\end{tabular}	
	}	
	\label{table:workload}
\end{table}

We run experiments on both bare-metal machines and small-scale clusters.
In addition, 
the evaluation is conducted on NSF Cloudlab~\cite{cloudlab}, which is hosted at the University of Utah, Downtown Data Center. 
Specifically, we use M510 as our physical machines for both standalone and cluster experiments. It contains
8-core Intel Xeon D-1548 at 2.0 GHz, 64GB ECC Memory and 256 GB NVMe flash storage. 

To evaluate the platforms from different perspectives, we conduct the following types of experiments.

\begin{itemize}

\item Type-1: 1 job in 1 container/pod (1 in 1): this set of experiments focuses on how the resource is allocated for limited workload within a container.

\item Type-2: 5 jobs in 1 container/pod (5 in 1): this set of experiments investigates the resource allocation for the same container with an increased workload.

\item Type-3: 5 jobs in 5 containers/pods (5 in 5): this set of experiments evaluate the system performance with a slightly larger number of individual containers.

\item Type-4: 5 jobs in 5 container/pods with \replaced[id=yuqi]{100s submission interval in Deep learning jobs and 20s interval for Hibench big data jobs}{100s submission interval}: this type of experiment introduces overlaps between computing jobs, which leads to more scheduling tasks for the system.

\item Type-5: Mix number of  different jobs with a random submission schedule: this type of experiment requires more scheduling tasks to be performed by the system.

\end{itemize}


When evaluating the systems, we conduct an experiment with and without resource limits.
Whenever the last job in the system is done, we terminate the containers in the system.
Throughout the experiments, we focus on the two key metrics to assess the platforms, completion time, and resource usage.

\begin{itemize}
\item Completion Time: the completion time of each individual job. Note that each container runs one or multiple jobs and each pod consists of one or multiple containers.

\item Resource Usage: resource usage when achieving the corresponding completion time per container/pod level. 

\end{itemize}

Additionally, we conduct experiments with the following job submission schedules.

\begin{itemize}
\item Burst schedule: it simulates a server with simultaneous workloads, which is a challenge for the system to adjust resources according to each individual job.

\item Fixed schedule: the time to launch a job is controlled by the administrator. When a new job joins the system, the system will have to redistribute the resources to accommodate it.

\item Random schedule: the launch times are randomized to simulate random submissions of jobs by users in a real cluster. The system has to frequently adjust resources to achieve its designed objectives.

\end{itemize}

\subsection{Evaluation Results}

The evaluation result is organized into  three parts in terms of different job categories, deep learning jobs, 
big data processing jobs, and a cluster with mixed settings.

\subsubsection{Deep Learning Jobs}
\label{sec:dl}

\begin{figure}[ht]
\centering
\includegraphics[width=1\linewidth]{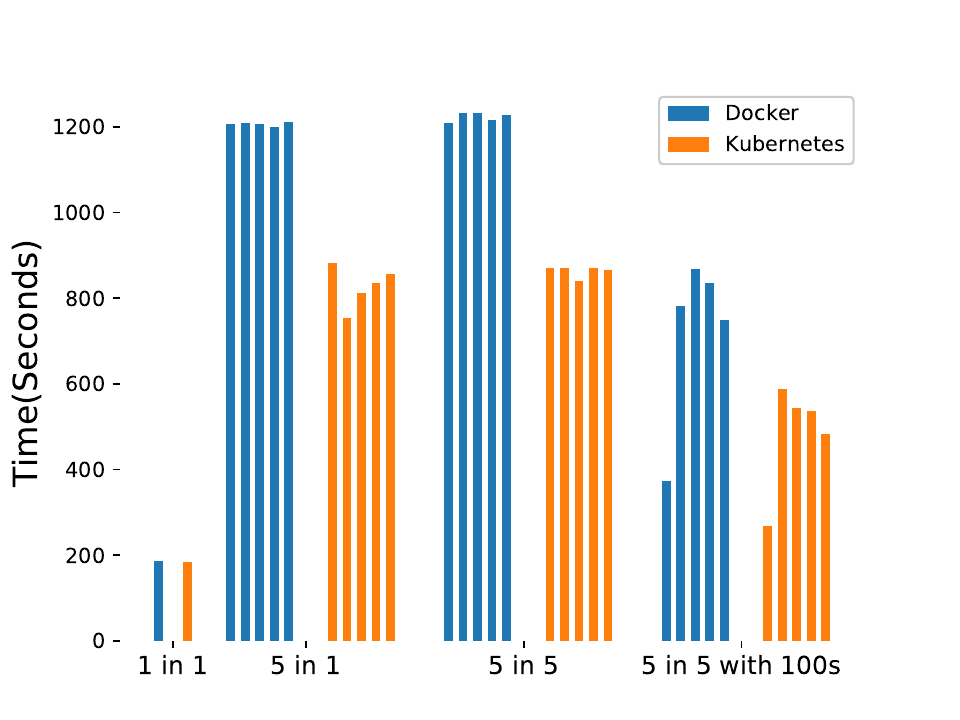}
\caption{Completion time of deep learning jobs}
\label{ct_dl} 
\end{figure}

Firstly, we conduct the above-mentioned Type 1 to 4 with deep learning jobs. We use Variational Autoencoders (VAE) on Tensorflow
as a representative workload. 
We utilize the default Tensorflow configuration for the experiments. Since deep learning applications are computation-intensive, we focus on the CPU usage in this subsection.


\begin{figure*}[ht]
   \centering
         \begin{subfigure}[b]{0.24\textwidth}
\centering
         \includegraphics[width=\textwidth]{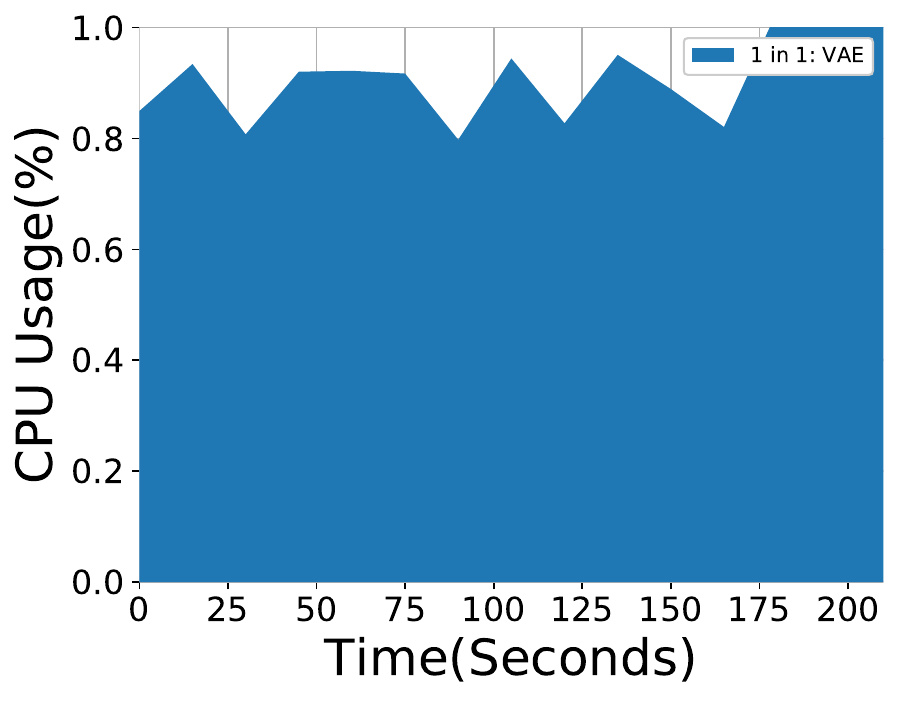}
\caption{1 in 1 VAE}
      \label{fig:DL:docker:1in1}
      \end{subfigure} 
      \begin{subfigure}[b]{0.24\textwidth}
\centering
         \includegraphics[width=\textwidth]{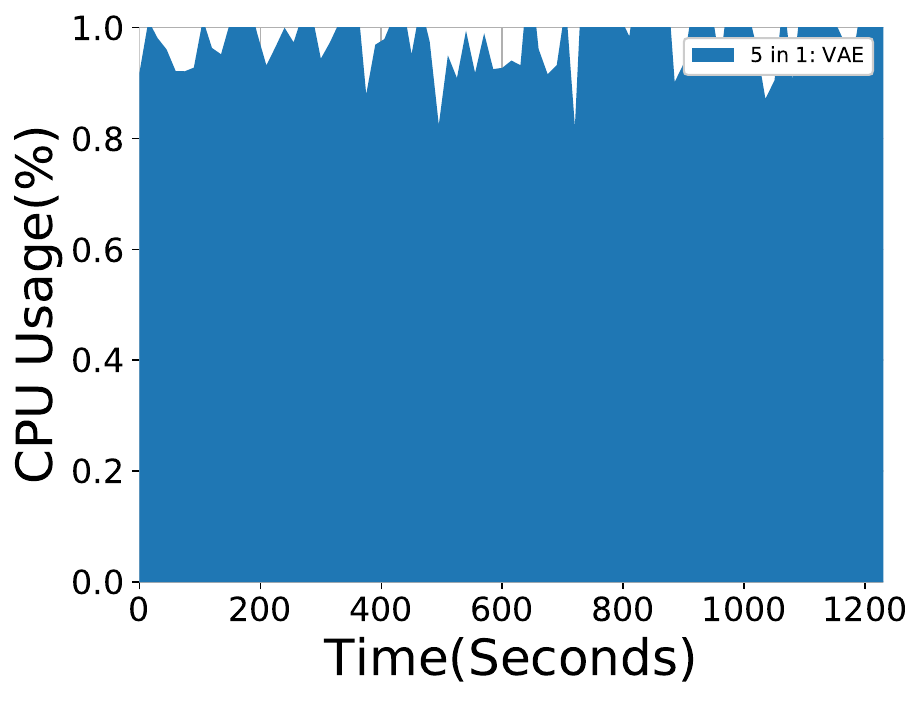}
\caption{5 in 1 VAE}
      \label{fig:DL:docker:5in1}
      \end{subfigure} %
      \begin{subfigure}[b]{0.24\textwidth}
\centering
         \includegraphics[width=\textwidth]{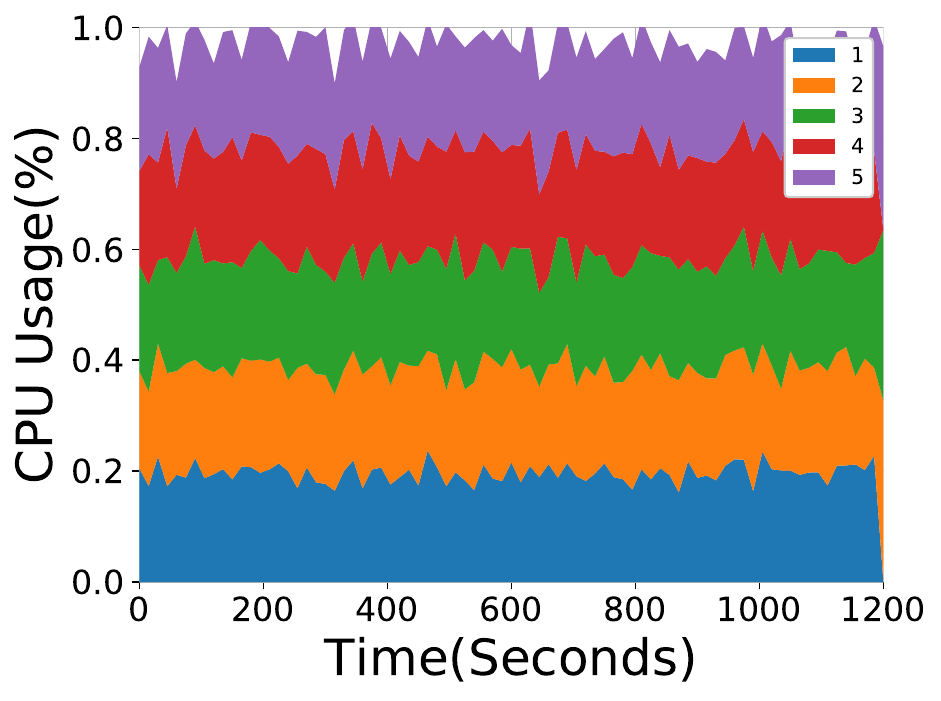}
\caption{5 in 5 VAE}
      \label{fig:DL:docker:5in5}
      \end{subfigure} %
      \begin{subfigure}[b]{0.24\textwidth}
	\centering
         \includegraphics[width=\textwidth]{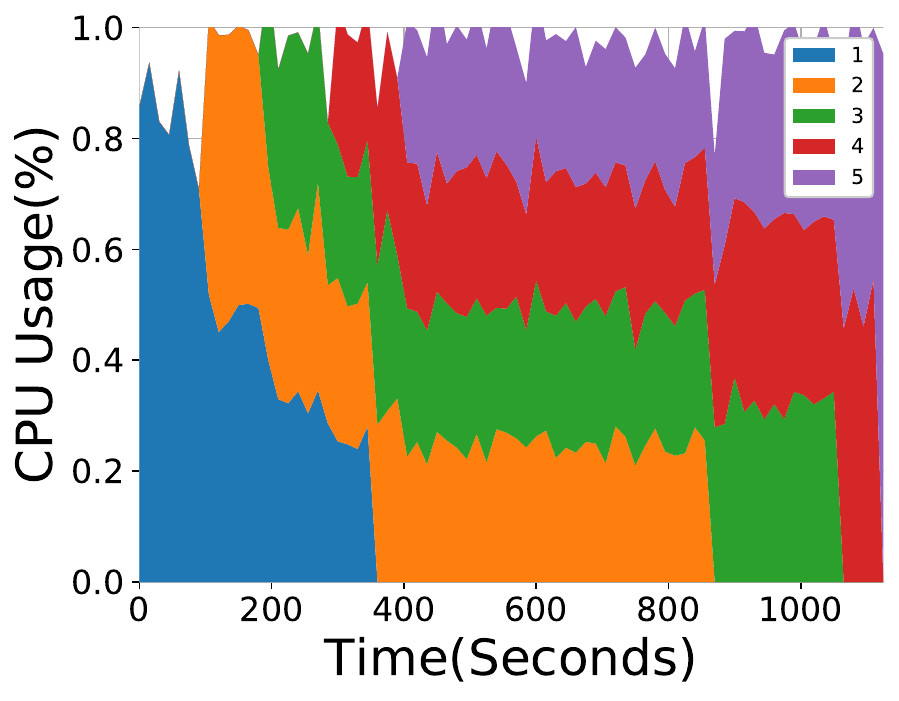}
	\caption{5 in 5 VAE with 100s interval}
      \label{fig:DL:docker:5in5:gap}
      \end{subfigure} %
\caption{CPU usage for VAE with Docker Platform}  
\label{docker:dl}               
\end{figure*}

\begin{figure*}[ht]
   \centering
         \begin{subfigure}[b]{0.24\textwidth}
\centering
         \includegraphics[width=\textwidth]{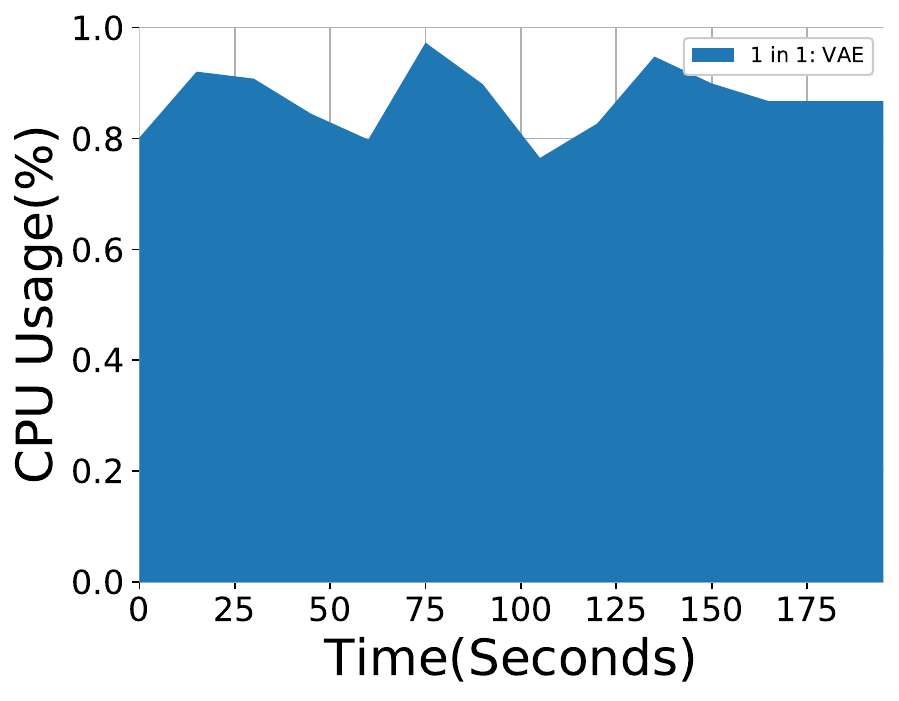}
\caption{1 in 1 VAE}
      \label{fig:DL:k8s:1in1}
      \end{subfigure} 
      \begin{subfigure}[b]{0.24\textwidth}
\centering
         \includegraphics[width=\textwidth]{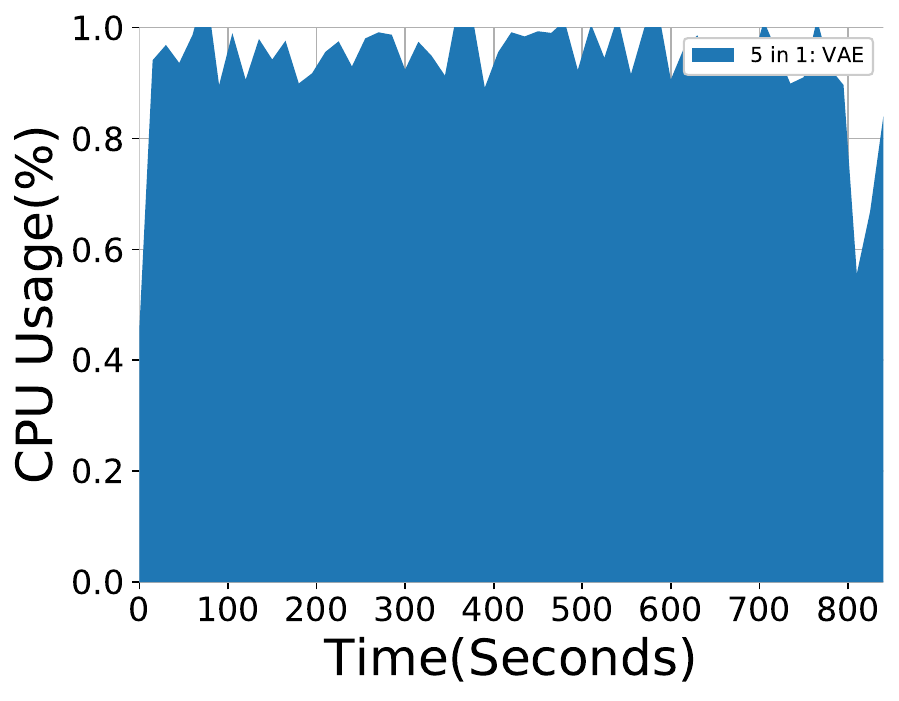}
\caption{5 in 1 VAE}
      \label{fig:DL:k8s:5in1}
      \end{subfigure} %
      \begin{subfigure}[b]{0.24\textwidth}
\centering
         \includegraphics[width=\textwidth]{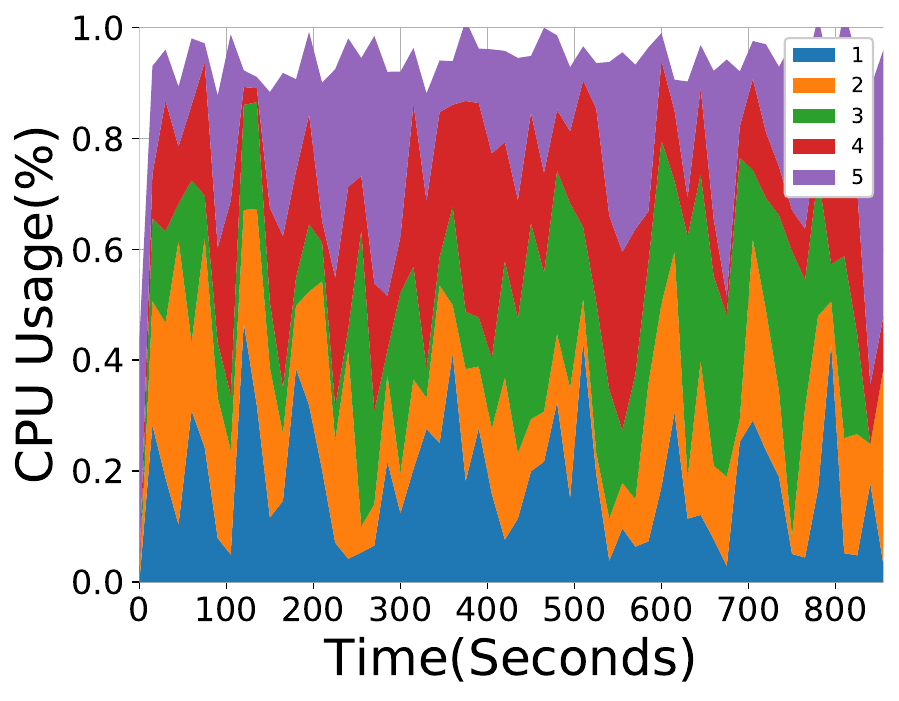}
\caption{5 in 5 VAE}
      \label{fig:DL:k8s:5in5}
      \end{subfigure} %
      \begin{subfigure}[b]{0.24\textwidth}
	\centering
         \includegraphics[width=\textwidth]{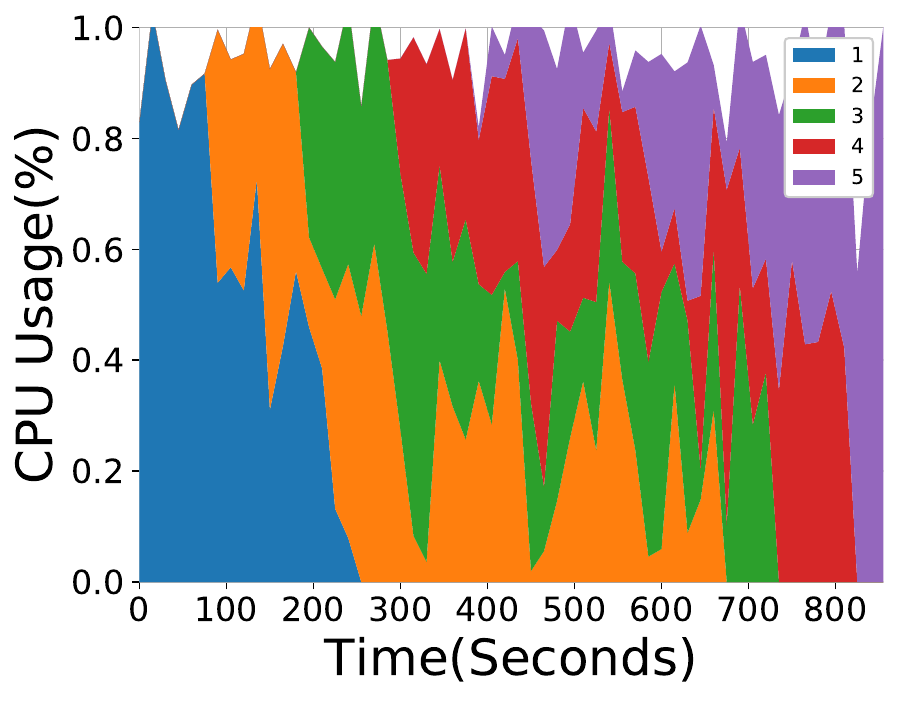}
	\caption{5 in 5 VAE with 100s interval}
      \label{fig:DL:k8s:5in5:gap}
      \end{subfigure} %
\caption{CPU usage for VAE with Kubernetes Platform}  
\label{k8s:dl}               
\end{figure*}

{\bf Completion Time}: Figure~\ref{ct_dl} plots the completion time of each type of experiment.
From the figure, we can see that if there is only 1 job running in the system, both Docker and Kubernetes
achieve similar results \replaced[id=yuqi]{(186.1s v.s. 183.3s)}{(278.9s v.s. 283.2s)}. 
When the number of containers/pods increases, the performance of Docker and Kubernetes varies a lot. 
Comparing the platforms with Type 2 experiments, where 5 VAE training jobs are running in 1 container, 
Kubernetes records a 37.7\% reduction on Job-2, which finishes at 753.2s and the same job of Docker completes at 1209.6s. 
The average completion time of 5 jobs is 1207.1s and 815.8s, an average 32.5\% reduction among 5 jobs.
Additionally, the standard deviation for the experiments is 4.47 and 38.79 for Docker and Kubernetes, respectively.
It indicates that, for the same amount of resources, Kubernetes achieve far better performance 
with intensive resource competition.

A similar trend is found with Type 3 experiment such that 5 VAE models are running concurrently in 5 containers. For example, the largest gain is discovered on Job-3, which 
reduces from 1233.7s to 840.8s, 31.87\%. On average, the improvement of completion time is 29.4\%, from 1223.8s to 863.6s. 
The standard derivations for Docker and Kubernetes are 8.12 and 12.02, respectively.

The same trend is observed again in the Type 4 experiments that 5 VAE jobs that run in their own container and are submitted to the system with 100s interval. The average completion time reduces from 772.2s to 482.6s for Docker and Kubernetes.
The largest gain is found at Job-3, who submits at 200s in the experiment and records 37.4\% reduction. 
The standard derivations are 178.84 (Docker) and 113.49 (Kubernetes). The values are significantly larger than all previous experiments since jobs are submitted into the system with an interval of 100s, which means Job-1 has 100s to occupy the entire
worker and Job-2 has 100s to share resources with only another one. The uneven resource distribution leads to different completion times.

When we compare experiment Type 3 with Type 2 experiments on the same platform, the completion time for both Docker and Kubernetes
increases. For example, the completion time of Job-2 in Docker and Kubernetes increases from 1209s to 1232s and 753s to 871s.
On average, it increases from 1207.1s to 1233.7s (Docker) and 815.8s to 863.6s.
This is because, with the same computing platform Docker or Kubernetes, when 5 jobs running in the 5 individual containers/pods, the overhead is larger than the same workload running in one single container/pod. More containers/pods lead to more repetitive works. 


Comparing to the two platforms, the system overhead can result in completion time variations,  however, it fails to 
explain the huge gaps in performance. 



{\bf Resource Usages}:
Next, we dig into the details of the previous experiments.
Figure~\ref{docker:dl} and Figure~\ref{k8s:dl} illustrate the CPU usage at the container (Docker) and pod (Kubernetes) level for the two platforms.

When only 1 job is running in the system, Figure~\ref{fig:DL:docker:1in1} and Figure~\ref{fig:DL:k8s:1in1} plot the data for Docker and Kubernetes, respectively.
We can see that the container in Docker uses a bit more CPU resources, 87\% on average than 83\% for Kubernetes.
This is due to the fact that Kubernetes utilizes an additional layer, pod, to manage a group of containers and thus, generate slightly more overhead with this 1 job workload.
With the same training job, however, Kubernetes is stopped at 195s and Docker terminates at 212s.
According to the completion time data, the job that runs inside each platform is 183.3s and 186.1s for Kubernetes and Docker.
Therefore, Docker requires more time to terminate a running container and completely release resources (26s vs 12s).

Next, we create a container/pod and submit 5 VAE jobs to the system.  Figure~\ref{fig:DL:docker:5in1} and Figure~\ref{fig:DL:k8s:5in1} present the CPU usage. We can find a similar trend that the container ins Docker consumes almost all the CPU usages and Kubernetes still reserves a small number of resources to the system itself. 

Considering completion time, while for Type 1 tests, both platforms achieve similar results, there is a huge performance gap when comparing them in Type 2, 3, 4 experiments, which fails to explain by system overhead.

Taking Fig.~\ref{fig:DL:docker:5in5} and Fig.~\ref{fig:DL:k8s:5in5} as an example, we find that, with the same workload and configuration, Docker and Kubernetes have very different resource management schemes.
In Fig.~\ref{fig:DL:docker:5in5} (Docker), each container is allocated an equal share of the resources, around 20\%, and maintains the same amount throughout the whole runtime. With Kubernetes (Fig.~\ref{fig:DL:k8s:5in5}), however, the system updates the resource allocation more frequently and there is no fixed limit to each of the containers. With this dynamic management scheme,
whenever an idling slot is found, Kubernetes can assign jobs without maintaining fairness. 
Given the same amount of resources, it can reduce the idling time and improve resource utilization, which leads to the reduction 
of completion time.
The similar results are found on Fig.~\ref{fig:DL:docker:5in5:gap} and Fig.~\ref{fig:DL:k8s:5in5:gap}, where there is a 100s submission gap between jobs.
The gap between jobs introduces more scheduling workload for the system.
We still can find the trends that are similar to before. With Docker the resource allocation is stable and Kubernetes manages the running pods with a more dynamic fashion. 
On one hand, Docker keeps maintaining fair resource allocation, e.g. from 100s to 200s, Job-1 and Job-2 average CPU usages are around 46\% each, 
on the other hand, Kubernetes dynamically adjusts resources to prevent idling, e.g. average CPU usages from 
100s to 200s are 51\% and 42\% for Job-1 and Job-2, respectively. The dynamic resource adjustment leads to performance improvement.



\begin{figure}
\centering
	\begin{subfigure}[t]{0.49\linewidth}
		\centering
		\includegraphics[width=1\textwidth]{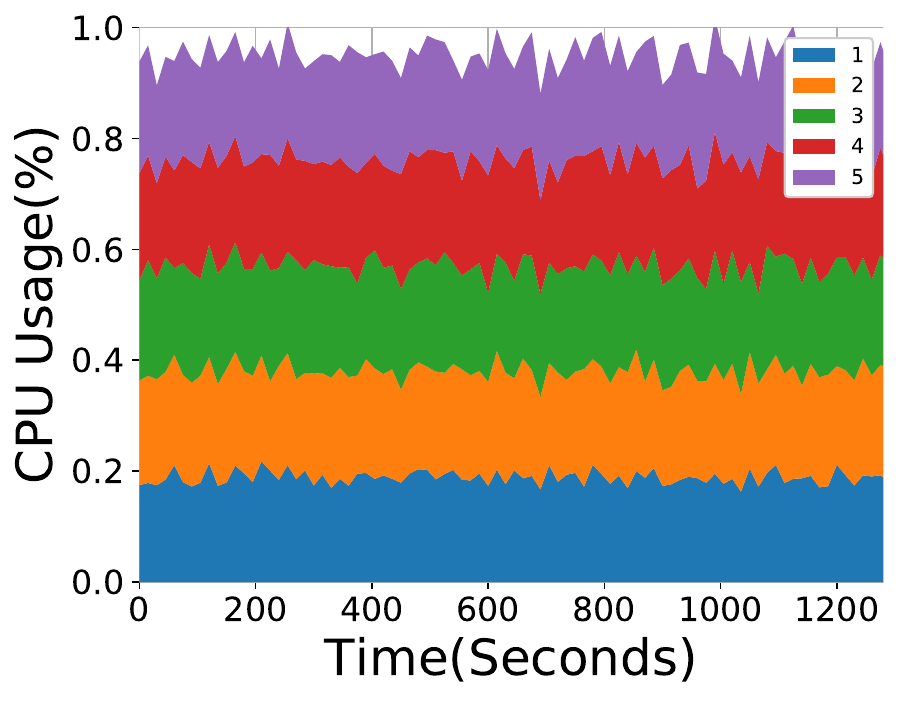}
		\caption{Docker}
		\label{fig:DL:docker:limited} 
	\end{subfigure}	%
	\begin{subfigure}[t]{0.49\linewidth}
	\centering
	\includegraphics[width=1\textwidth]{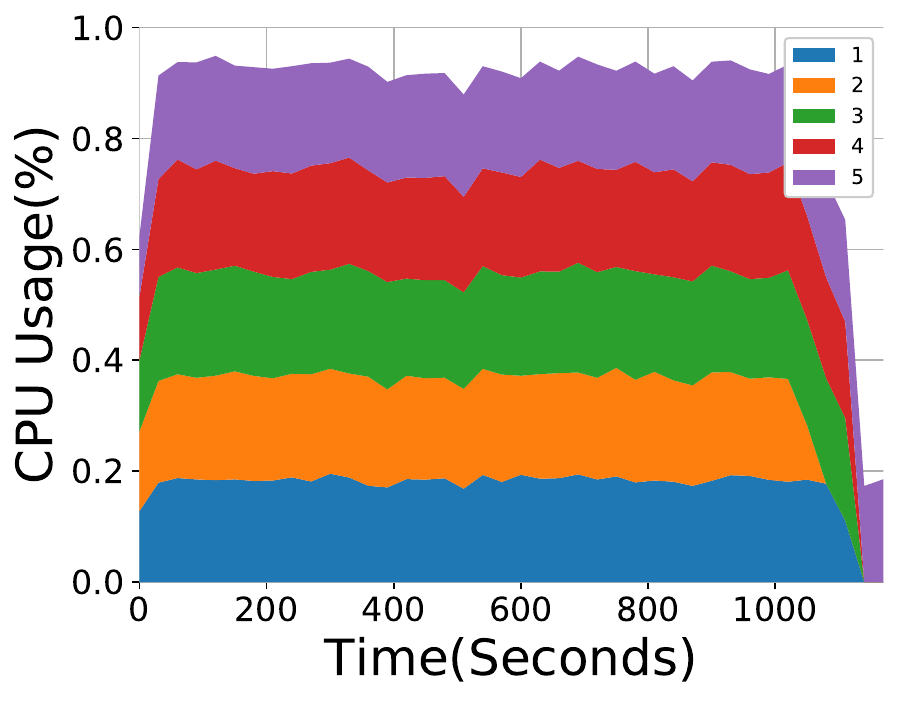}
	\caption{Kubernetes}
	\label{fig:DL:k8s:limited} 
	\end{subfigure}	
\caption{Experiments of 5 in 5 with specific limits}		
\end{figure}

{\bf Explicit Resource Limits}: 
Then, we re-conduct Type 3 experiments with an explicit limit for each of the container/pod to understand how resource configuration affects the performance. Please note that this value is an upper bound of the resources that can be occupied by 
a specific container.   
The similar of usage patterns are indicated on Figure~\ref{fig:DL:docker:5in5} and Figure~\ref{fig:DL:docker:limited}.
However, with a specific limit on each container, the system scheduler  
Comparing Figure~\ref{fig:DL:k8s:5in5} and Figure~\ref{fig:DL:k8s:limited}, a very different resource management behavior
is observed.  
We can see that the resource usage of each container is capped at 20\% of the total amount. With limits, the system lost 
the ability to dynamically adjust resource allocation. Therefore, we discover the performance downgrade when comparing two experiments on the same platform. Table~\ref{table:ct} presents the completion time of Docker and Kubernetes with default setting (D) and specific limits setting (L). 
Comparing Docker (D) with Docker (L), the completion time increases around 1\% to 6\%. The reason lies in the fact that,
with specific limit values on each container, it introduces more scheduling overhead for the system to keep tracking the resource
consumption. When considering Kubernetes, not only the scheduling overhead is brought by maintaining the limits, but also the explicit limits disable the flexibility for scheduling to assign tasks to maximize 
resource utilization and preventing idling tasks. As we can see that Job-5 records completion time increase of 38.5\% and on average, it increases from 863.6s to 1128.2s, 30.6\%.


\begin{table}[ht]\small
	\centering
	\caption{Experiments with Specific Limits (Seconds)}
	\scalebox{0.95}{
		
		\begin{tabular}{ | c | c | c | c | c | c |}				
			\hline
			Platform    & Job-1& Job-2 & Job-3 & Job-4 & Job-5  \\ \hline	
			Docker (D) & 1209.1 & 1232.5 & 1233.7 & 1216.4 & 1229.1  \\ \hline				
			Docker (L) & 1277.6 & 1270.5 & 1255.3 & 1278.2 & 1290.3  \\ \hline	
			Kubernetes (D) & 871.3 & 872.1 & 840.8 & 871.9 & 867.2 \\ \hline
			Kubernetes (L) & 1124.7 & 1056.6 & 1125.7 & 1138.5 & 1198.4 \\ \hline
						
		\end{tabular}	
	}	
	\label{table:ct}
\end{table}


{\bf Parallelism with Multithreading}
Finally, we study how multithreading configuration can affect the performance.
Generally, developers take advantage of multithreading to perform several tasks simultaneously
and maximize multi-core hardware. However, incorrect usage of multithreading may result in increased hardware usage cycles and can drastically reduce the system performance.

In these experiments, we use Pytorch as our computing framework. In Pytorch, developers can use a method \texttt{torch.set_num_threads(int)} to set the number of threads used for intraop parallelism.
The default value is 8, which means that each Pytorch task can launch 8 threads for parallel processing.
\begin{figure}[!htb]
\centering
\includegraphics[width=1\linewidth]{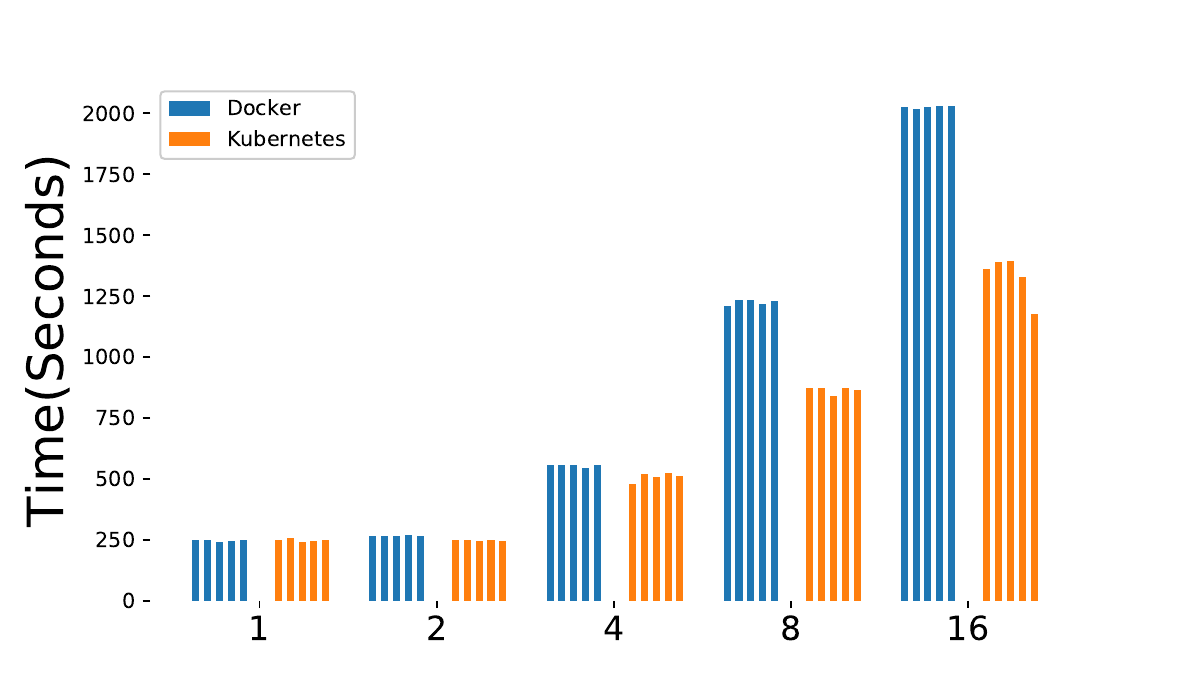}
\caption{Completion time with various thread numbers}
\label{fig:thread} 
\end{figure}
\begin{figure*}[!htb]
   \centering
         \begin{subfigure}[b]{0.30\textwidth}
\centering
         \includegraphics[width=\textwidth]{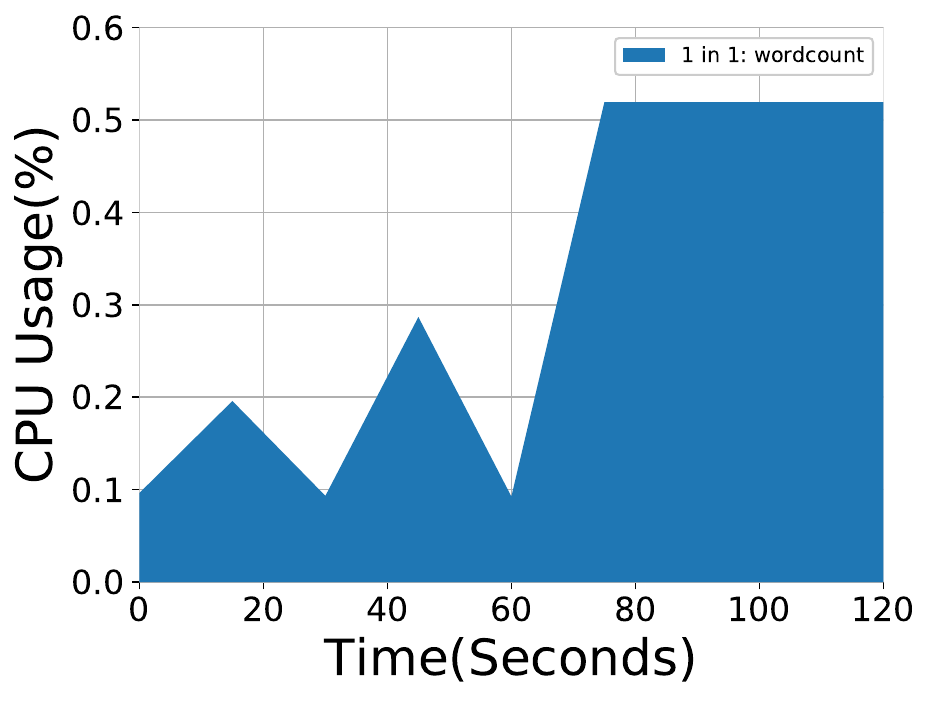}
\caption{1 in 1 WordCount}
      \label{fig:BD:docker:wc:1in1}
      \end{subfigure} 
      \begin{subfigure}[b]{0.30\textwidth}
\centering
         \includegraphics[width=\textwidth]{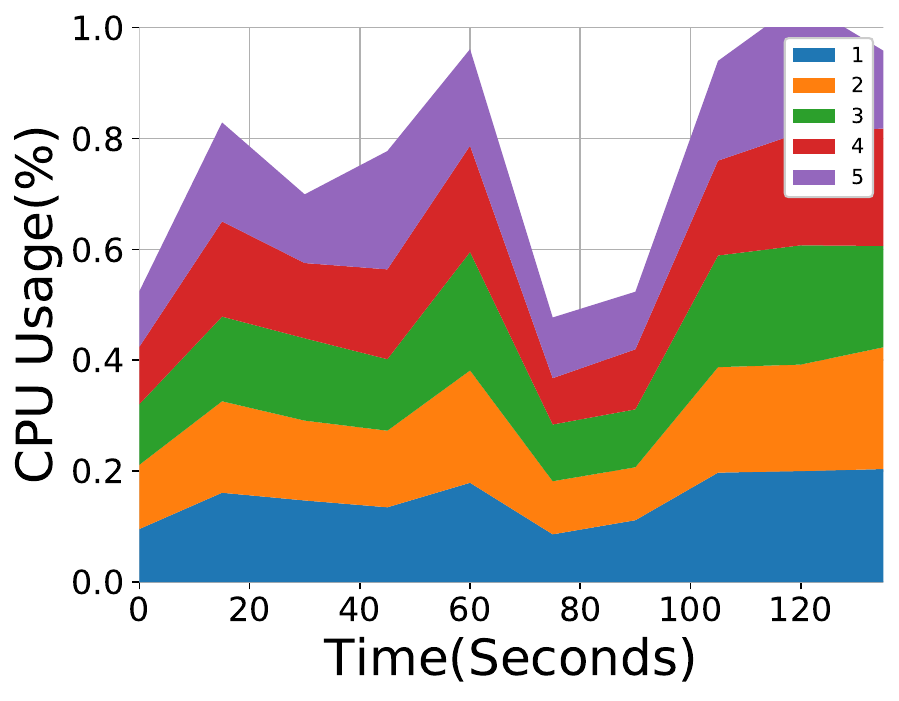}
\caption{5 in 5 WordCount}
      \label{fig:BD:docker:wc:5in5}
      \end{subfigure} %
      \begin{subfigure}[b]{0.30\textwidth}
\centering
         \includegraphics[width=\textwidth]{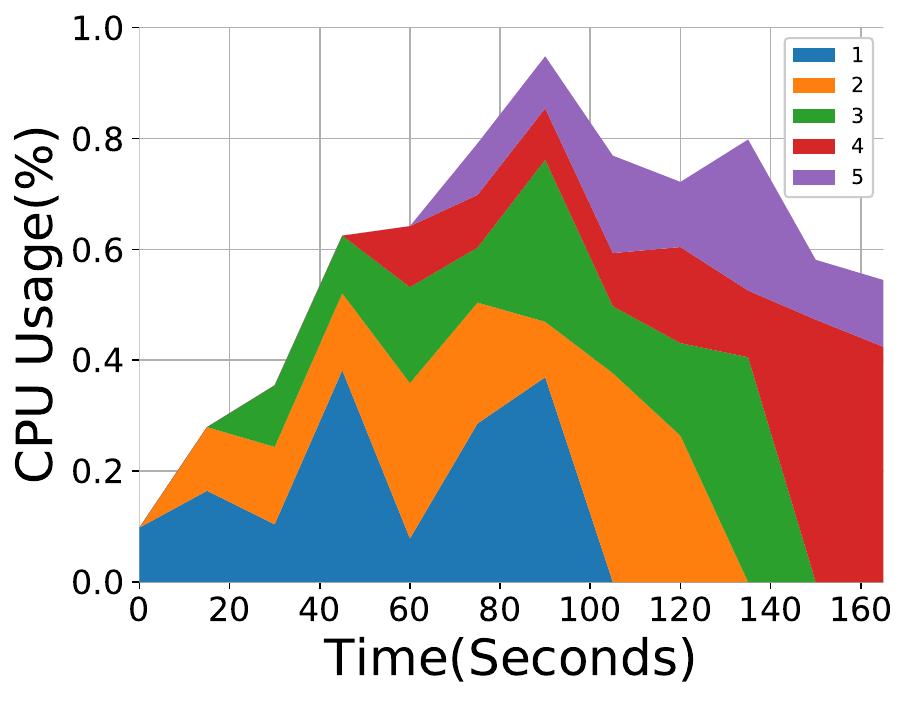}
\caption{5 in 5 WordCount with 100s interval}
      \label{fig:BD:docker:wc:5in5:gap}
      \end{subfigure} %
\caption{CPU usage for Hadoop Yarn (WordCount) with Docker Platform}        
\label{docker:wc}               
\end{figure*}

\begin{figure*}[!htb]
   \centering
         \begin{subfigure}[b]{0.30\textwidth}
\centering
         \includegraphics[width=\textwidth]{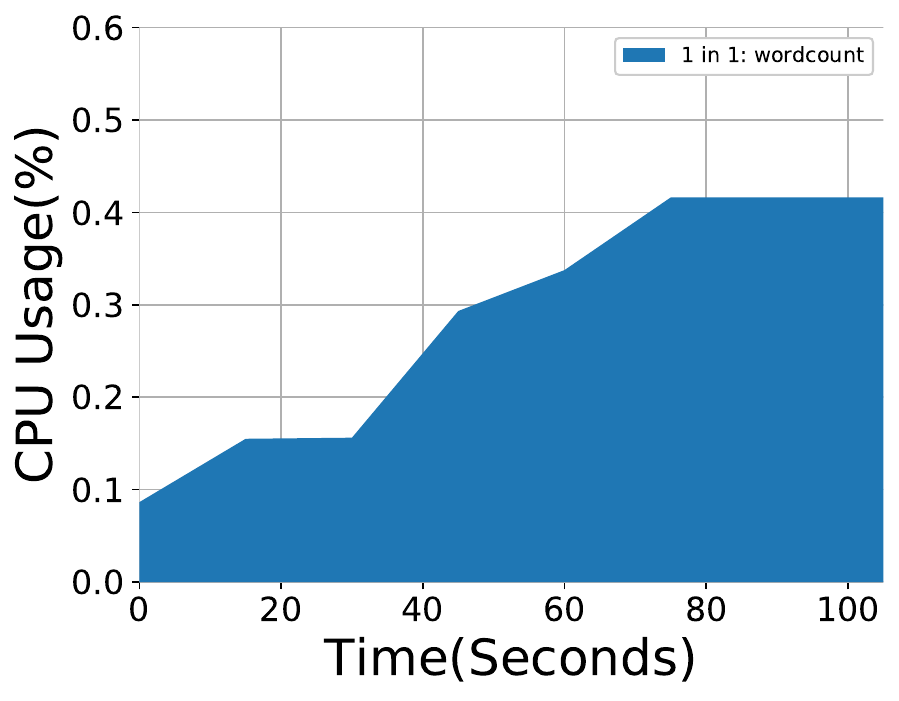}
\caption{1 in 1 WordCount}
      \label{fig:BD:k8s:wc:1in1}
      \end{subfigure} 
      \begin{subfigure}[b]{0.30\textwidth}
\centering
         \includegraphics[width=\textwidth]{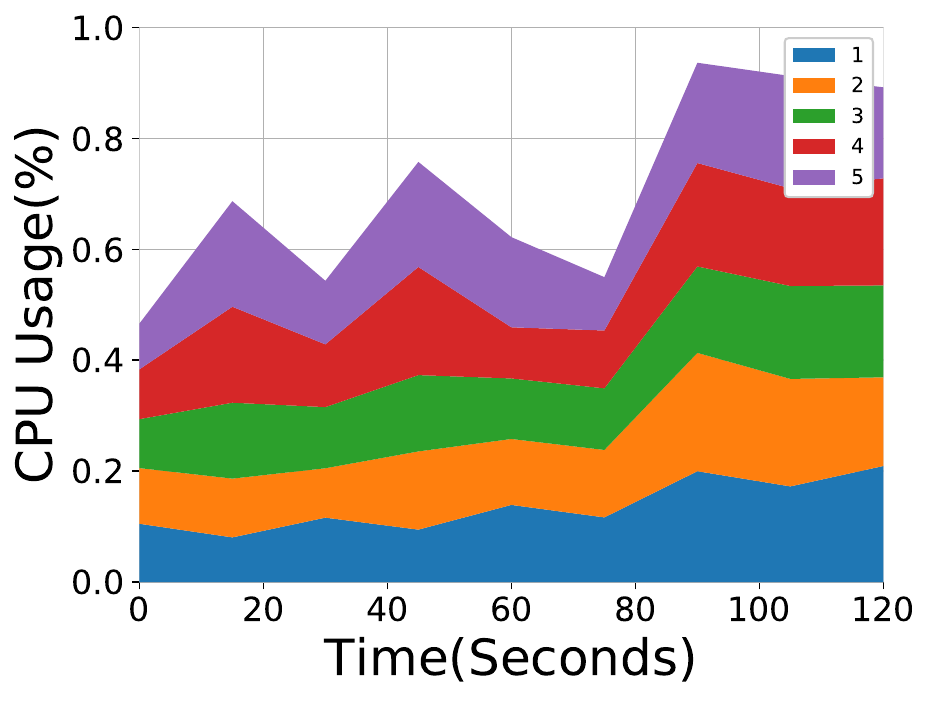}
\caption{5 in 5 WordCount}
      \label{fig:BD:k8s:wc:5in5}
      \end{subfigure} %
      \begin{subfigure}[b]{0.30\textwidth}
\centering
         \includegraphics[width=\textwidth]{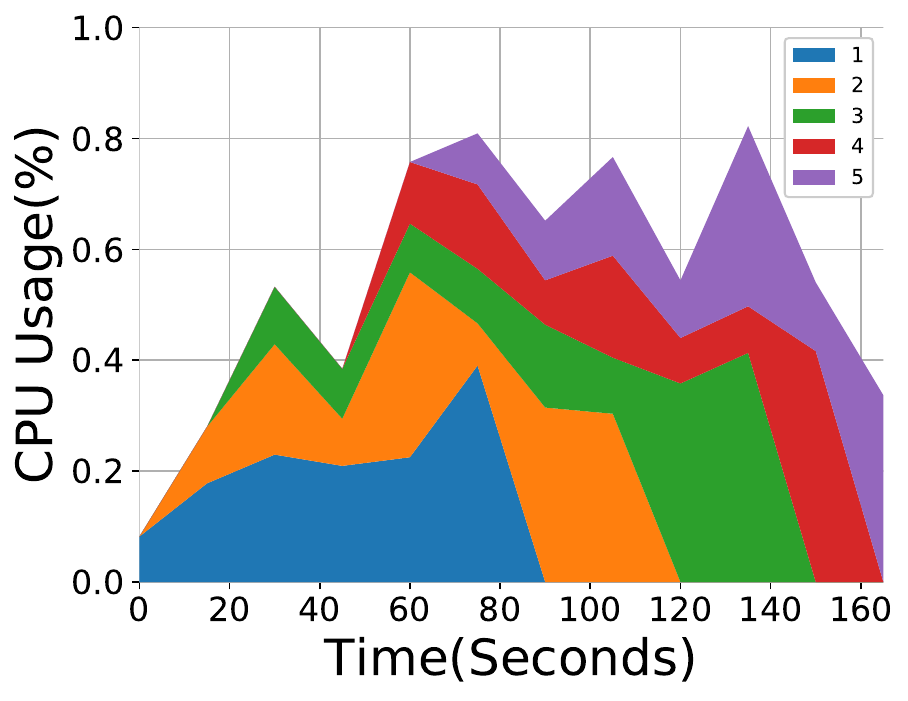}
\caption{5 in 5 WordCount with 100s interval}
      \label{fig:BD:k8s:wc:5in5:gap}
      \end{subfigure} %
\caption{CPU usage for Hadoop Yarn (WordCount) with Kubernetes Platform}               
\label{k8s:wc}               
\end{figure*}

Fig.~\ref{fig:thread} illustrates the completion time under different thread settings.
Obviously, the performance varies a lot. 
With the default setting 8 threads, the average completion time is 1207.1s and 815.8s for Docker and Kubernetes, respectively.
When increasing the value to 16, the average becomes 2025.8s and 1329.4s, which increases 67.8\% and 62.9\%.
The significant performance downgrade is due to the increased overhead for switching between threads.
When reducing the number to 4, the values reduce to 554.2s and 508.6s for Docker and Kubernetes.
The average CPU usage for all three settings is at around 97\% in average. Given the same resource utilization, 
a smaller number of threads results in less swapping overhead and lead to better performance.
Further reducing the thread number to 1 and 2, they achieve similar completion time 249.0s, 253.9s (Docker) and
248.4s, 249.6s (Kubernetes). Both of them utilize fewer resources than the total availability. This implies that
the performance in these experiments mainly depends on the algorithm implementation, not the resources.  

Please note that optimal value that can achieve the best performance varies in different systems and workloads. It mainly depends on the details of hardware type, configuration, model characteristics, and algorithm implementation.

\subsubsection{Big Data Processing}


In this subsection, we evaluate the systems with Hadoop Yarn and Spark. 

With WordCount (1 Map 1 Reduce) as the representative workload, we conduct three types of experiments 1 in 1, 5 in 5 and 5 in 5 with \replaced[id=yuqi]{20s}{100s} submission interval.
Figure~\ref{docker:wc} and Figure~\ref{k8s:wc} present the results of Hadoop Yarn. 
When there is 1 job running in the system, the resource contention is minimum. 
However, we can see from Figure~\ref{fig:BD:docker:wc:1in1}
and Figure~\ref{fig:BD:k8s:wc:1in1}, the container/pod fails to consume all the CPU resource (16 cores).
This indicates that the WordCount job is not as computation-intensive as deep learning applications. 
The resource usage between two platforms is similar except the pod in Kubernetes gets more dynamic allocation.

\begin{comment}
\begin{figure}[ht]
\centering
\includegraphics[width=1\linewidth]{figs/eval/docker_5in5_mem-eps-converted-to.pdf}
\caption{Memory usage for 5 in 5 Hadoop Yarn (WordCount) with Docker platform}
\label{fig:docker:wc:mem} 
\end{figure}

\begin{figure}[ht]
\centering
\includegraphics[width=1\linewidth]{figs/eval/k8s_5in5_mem_pod-eps-converted-to.pdf}
\caption{Memory usage for 5 in 5 Hadoop Yarn (WordCount) with Kubernetes platform}
\label{fig:k8s:wc:mem} 
\end{figure}
\end{comment}

\begin{figure}
\centering
	\begin{subfigure}[t]{0.49\linewidth}
		\centering
		\includegraphics[width=1\linewidth]{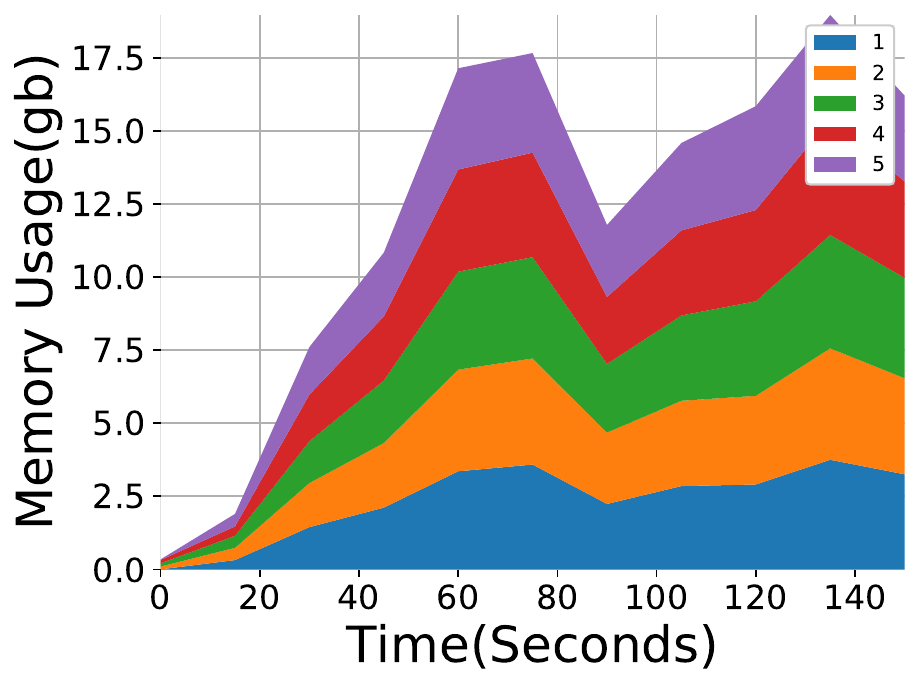}
		\caption{Docker}
		\label{fig:docker:wc:mem} 
	\end{subfigure}	%
	\begin{subfigure}[t]{0.49\linewidth}
		\centering
		\includegraphics[width=1\linewidth]{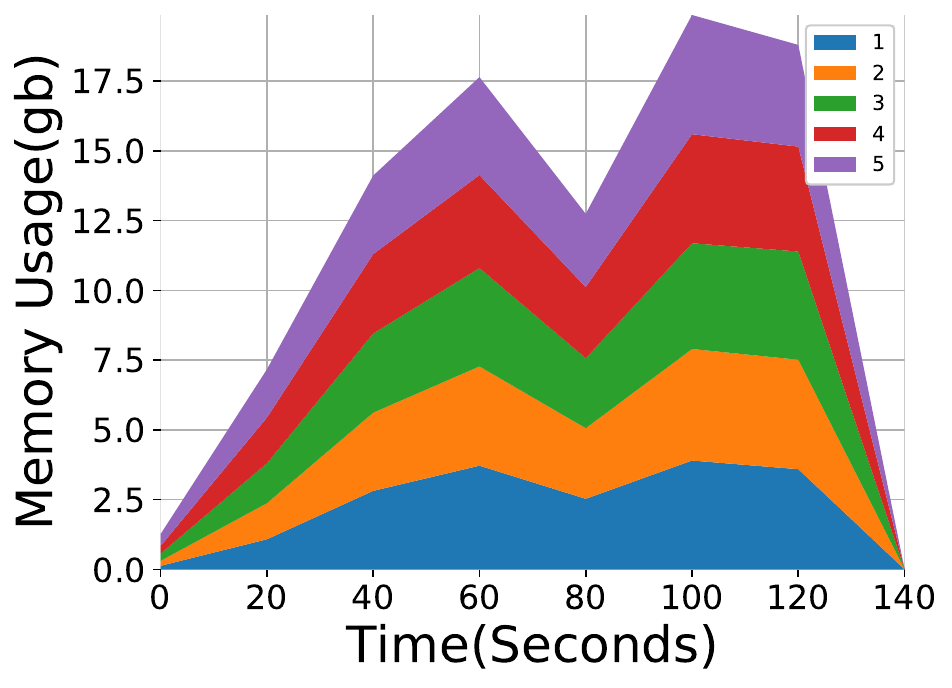}
		\caption{Kubernetes}
		\label{fig:k8s:wc:mem} 
	\end{subfigure}	
\caption{Memory usage of Hadoop Yarn with 5 in 5 (Type 3) WordCount with specific limits}		
\end{figure}

\begin{comment}
\begin{figure}
\centering
	\begin{subfigure}[t]{0.49\linewidth}
		\centering
        \includegraphics[width=1\linewidth]{figs/eval/ct_wc-eps-converted-to.pdf}
        \caption{WordCount}
        \label{ct_wc} 
	\end{subfigure}	%
	\begin{subfigure}[t]{0.49\linewidth}
		\centering
        \includegraphics[width=1\linewidth]{figs/eval/ct_pr-eps-converted-to.pdf}
        \caption{PageRank}
        \label{ct_pr}  
	\end{subfigure}	
\caption{Completion time for WordCount and PageRank jobs}		
\end{figure}
\end{comment}

\begin{figure}[ht]
\centering
\includegraphics[width=1\linewidth]{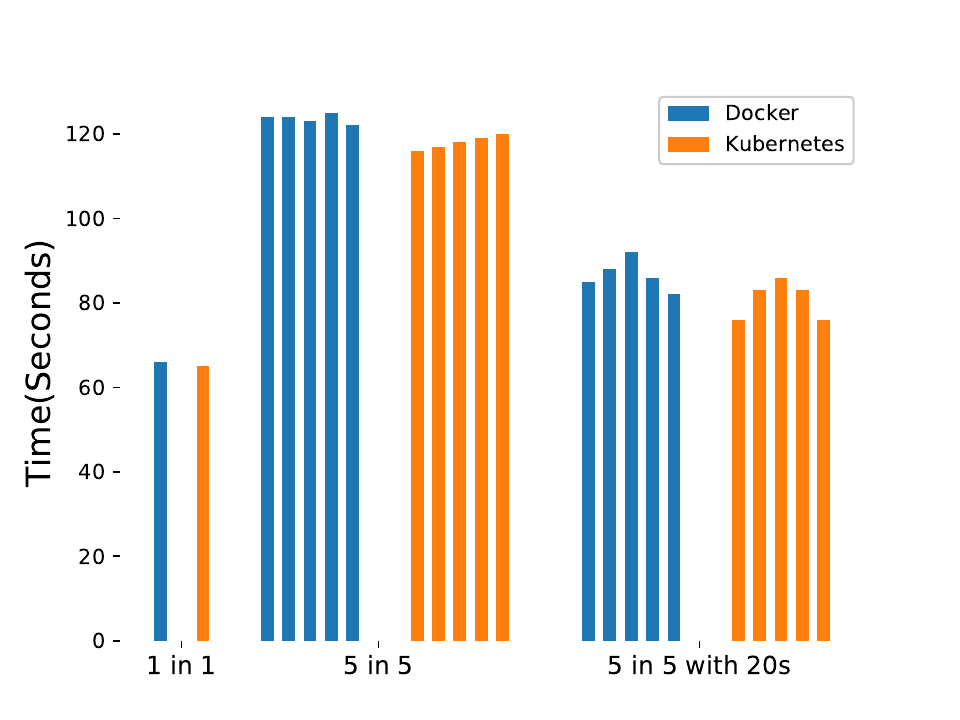}
\caption{Completion time of WordCount jobs}
\label{ct_wc} 
\end{figure}

\begin{figure}[ht]
\centering
\includegraphics[width=1\linewidth]{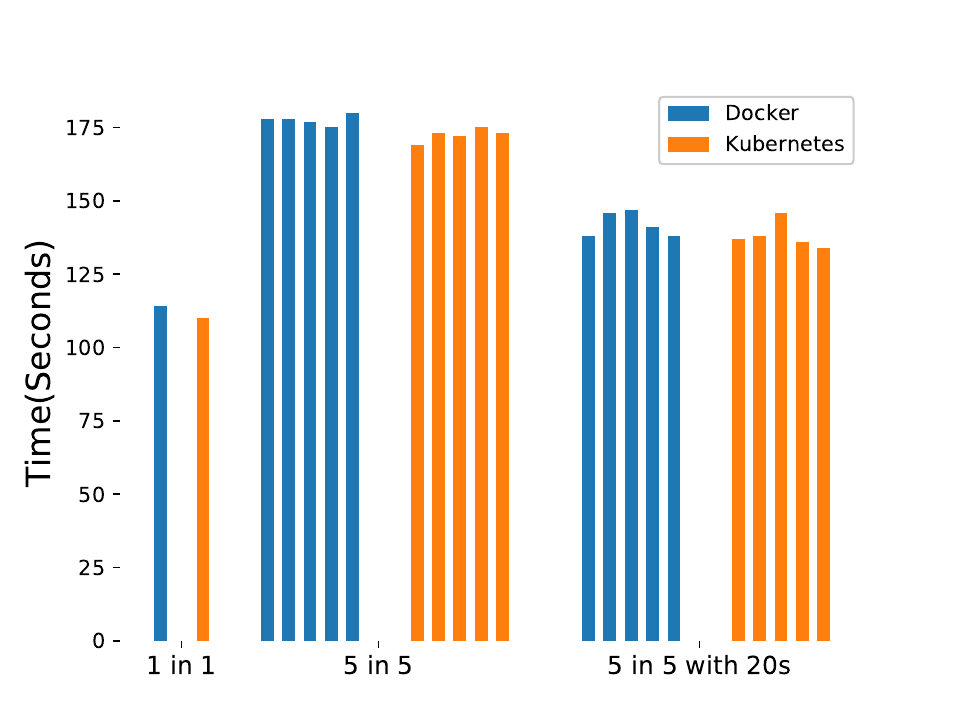}
\caption{Completion time of PageRank jobs}
\label{ct_pr} 
\end{figure}

 
Figure~\ref{fig:BD:docker:wc:5in5} and Figure~\ref{fig:BD:k8s:wc:5in5} show  the results when 5 WordCount jobs in 5 individual containers/pods. With the figures, we can clearly see the Map-Reduce transition phase, where the CPU usage decreases first as the map is finishing and then increase again as the reduction is running. 

When submitting a workload with a gap, more scheduling tasks need to be done by the system. 
Figure~\ref{fig:BD:docker:wc:5in5:gap} and Figure~\ref{fig:BD:k8s:wc:5in5:gap} plot the results.
Kubernetes can dynamically change resource allocation with respect to new workloads.

Besides CPU usage, Figure~\ref{fig:docker:wc:mem} and Figure~\ref{fig:k8s:wc:mem} illustrate the memory usage of 5 in 5 experiments. The trend of memory usage is very similar between the two platforms. However, the memory on Docker is released 
at time 530.6s and with Kubernetes, the memory is returned to the system at 680.8s, which is 28.3\% longer than Docker.

Figure~\ref{ct_wc} illustrate the completion time of the WordCount jobs. 
Unlike a huge performance difference with deep learning jobs, the completion time of WordCount jobs is within a small range.
\replaced[id=yuqi]{For example, the average completion time for 1 in 1, 5 in 5 and 5 in 5 with 100s submission interval is 66s v.s. 65s, 123.6s v.s. 118s, and 86.6s v.s. 80.6s.}
{For example, the average completion time for 1 in 1, 5 in 5 and 5 in 5 with 100s submission interval is 241.7s v.s. 245.0s, 554.1s v.s. 582.1s, and 289.8s v.s. 297.8s.} The reason is that WordCount jobs are far less CPU-sensitive than VAE. Additionally, MapReduce based WordCount jobs require diverse types of resources, memory for the data, and network for shuffling.

\begin{figure*}[ht]
   \centering
         \begin{subfigure}[b]{0.30\textwidth}
\centering
         \includegraphics[width=\textwidth]{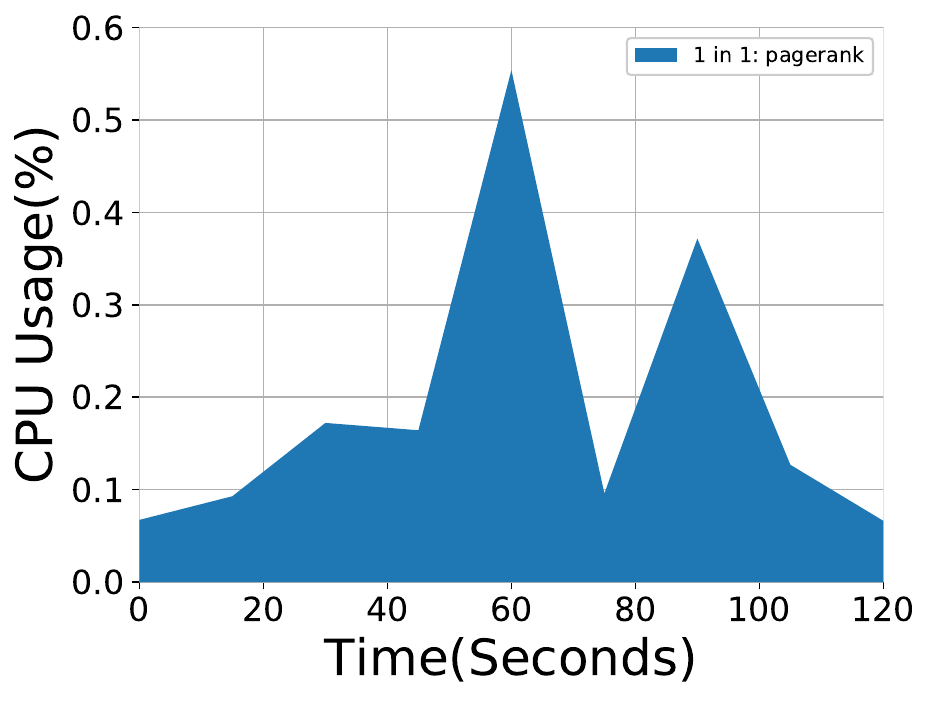}
\caption{1 in 1 PageRank}
      \label{fig:BD:docker:pr:1in1}
      \end{subfigure} 
      \begin{subfigure}[b]{0.30\textwidth}
\centering
         \includegraphics[width=\textwidth]{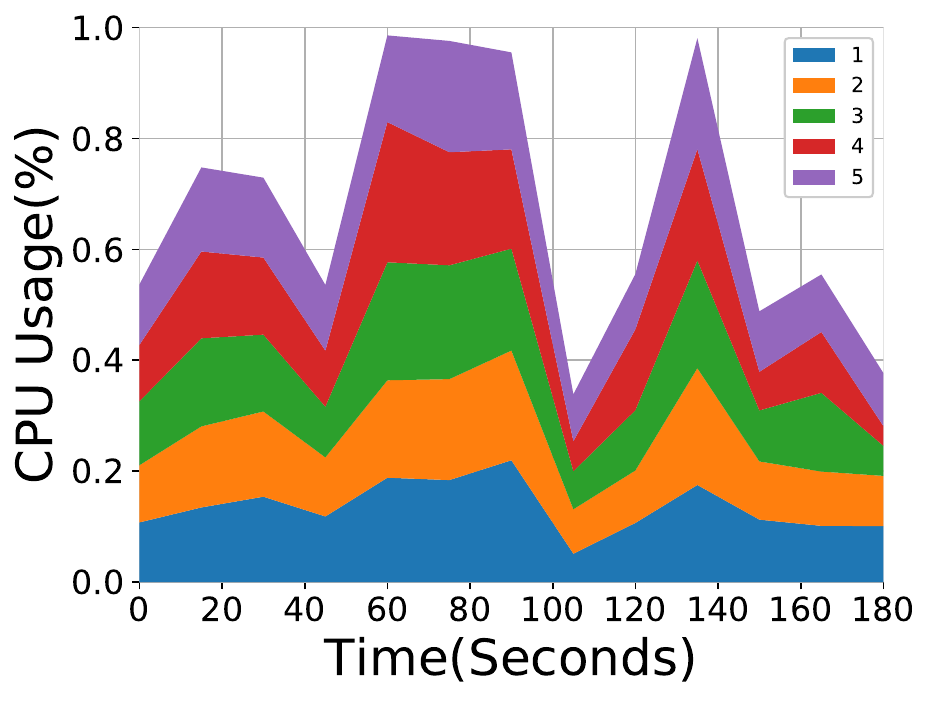}
\caption{5 in 5 PageRank}
      \label{fig:BD:docker:pr:5in5}
      \end{subfigure} %
      \begin{subfigure}[b]{0.30\textwidth}
\centering
         \includegraphics[width=\textwidth]{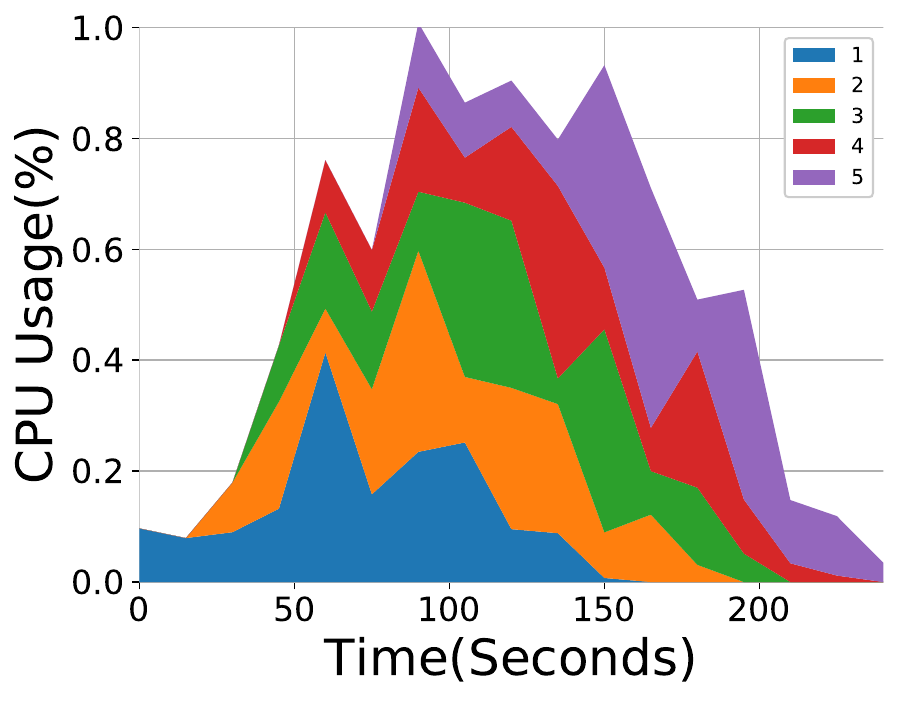}
\caption{5 in 5 PageRank with 100s interval}
      \label{fig:BD:docker:pr:5in5:gap}
      \end{subfigure} %
\caption{CPU usage for Spark (PageRank) with Docker}                     
\label{docker:pr}               
\end{figure*}

\begin{figure*}[ht]
   \centering
         \begin{subfigure}[b]{0.30\textwidth}
\centering
         \includegraphics[width=\textwidth]{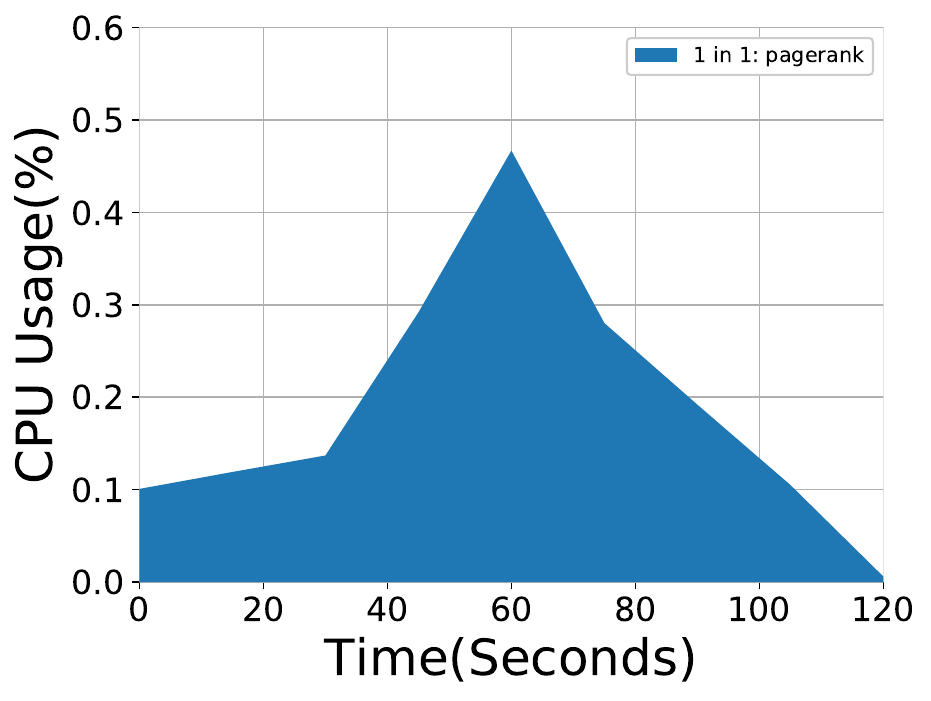}
\caption{1 in 1 PageRank}
      \label{fig:BD:k8s:pr:1in1}
      \end{subfigure} 
      \begin{subfigure}[b]{0.30\textwidth}
\centering
         \includegraphics[width=\textwidth]{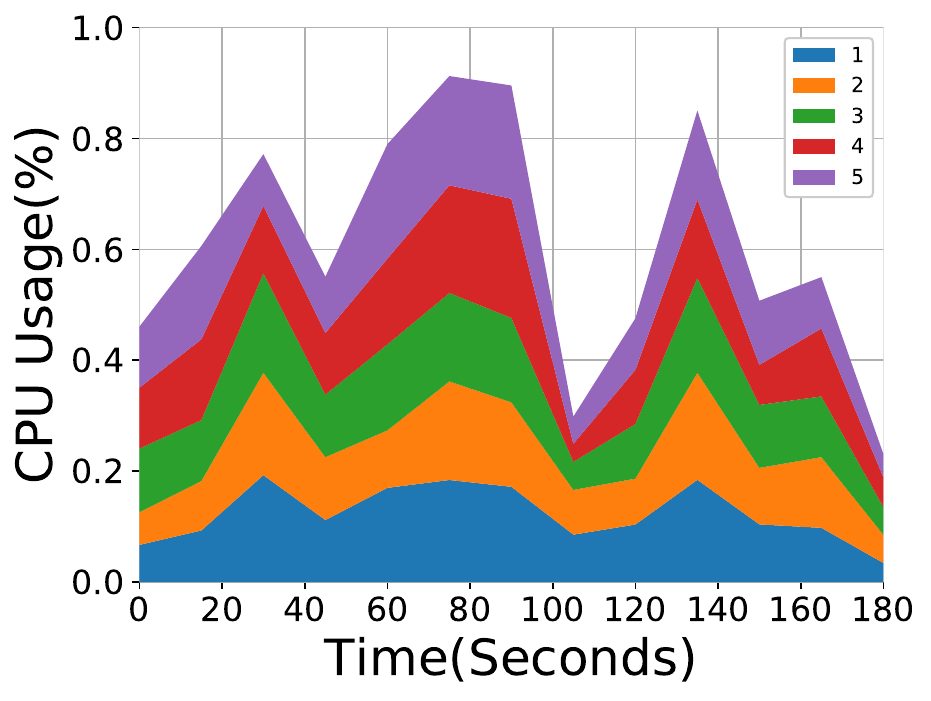}
\caption{5 in 5 PageRank}
      \label{fig:BD:k8s:pr:5in5}
      \end{subfigure} %
      \begin{subfigure}[b]{0.30\textwidth}
\centering
         \includegraphics[width=\textwidth]{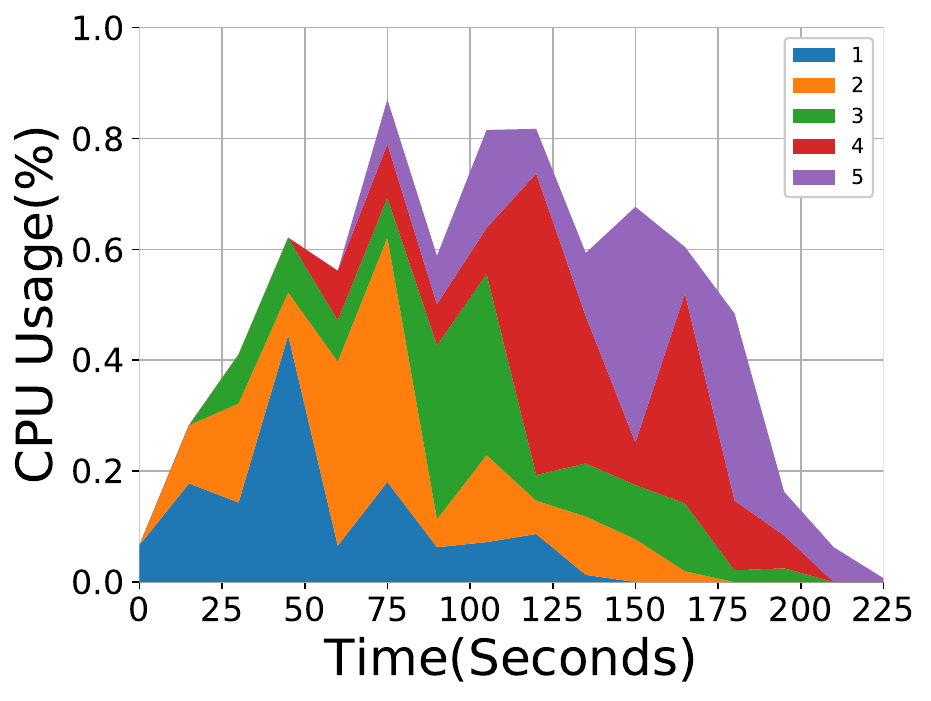}
\caption{5 in 5 PageRank with 100s interval}
      \label{fig:BD:k8s:pr:5in5:gap}
      \end{subfigure} %
\caption{CPU usage for Spark (PageRank) with Kubernetes}                     
\label{k8s:pr}               
\end{figure*}

Figure~\ref{docker:pr} and Figure~\ref{k8s:pr} present the results for PageRank jobs on Spark with 1 in 1, 5 in 5 and 5 in 5 with 100s submission interval.
We find multiple stages from the resource usage patterns, which is a feature of PageRank jobs. In general, jobs in Kubernetes consumes fewer resources since the system reserves part of the resources for itself.
Figure~\ref{ct_pr} plots the completion time of PageRank jobs. The same as WordCount jobs, we find that both Docker and Kubernetes achieve a similar completion time in the experiments.

\begin{figure*}[ht]
\centering
\includegraphics[width=1\linewidth]{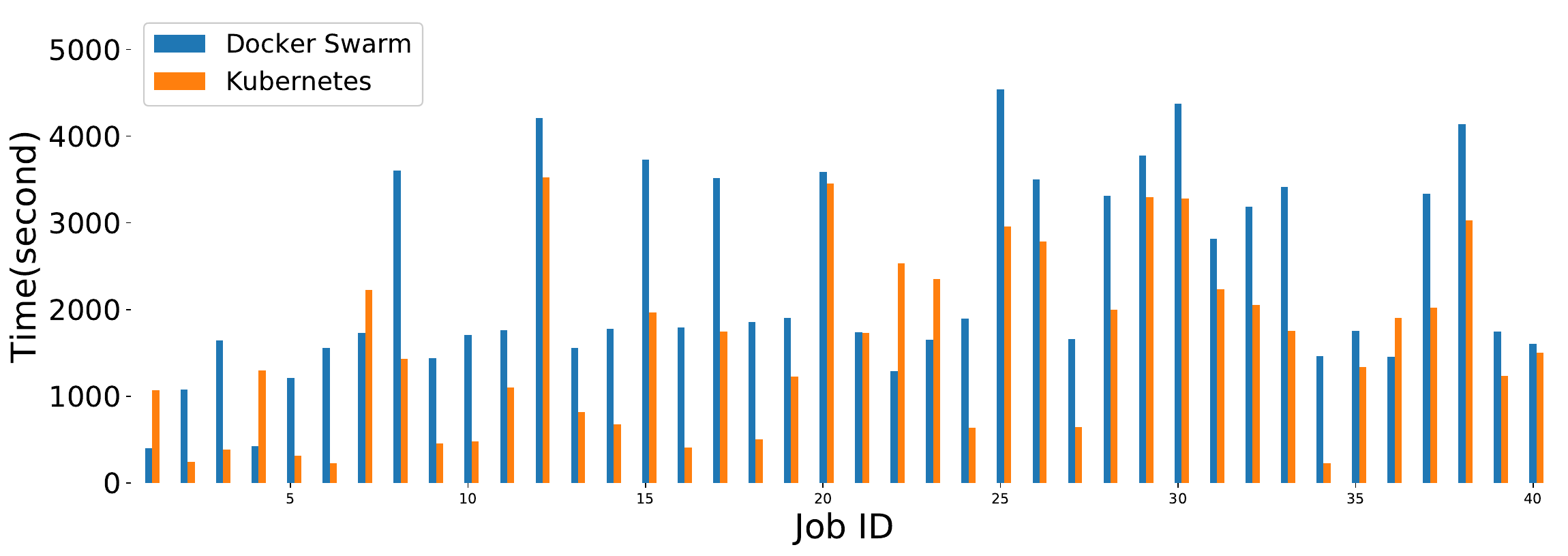}
\caption{Completion time of 40 deep learning jobs in a 4-node cluster}
\label{cluster_bar} 
\end{figure*}

\subsubsection{Cluster Environment} 
In this subsection, we evaluate the two platforms with a 4-node cluster, 1 manager, and 4 workers, which builds with previously mentioned M510 machines. In this cluster, Worker-1 serves as both manager and worker.

For the experiments, we first randomly select 40 jobs from deep learning applications in Table~\ref{table:workload}.
Then, for each job, we randomly generate a submission time with the interval [0, 1000s].
With this setting, we produce more tasks for the system scheduler, which is responsible for various tasks, such as container placement and resource allocation.

\begin{figure*}[h]
   \centering
         \begin{subfigure}[b]{0.24\textwidth}
\centering
         \includegraphics[width=\textwidth]{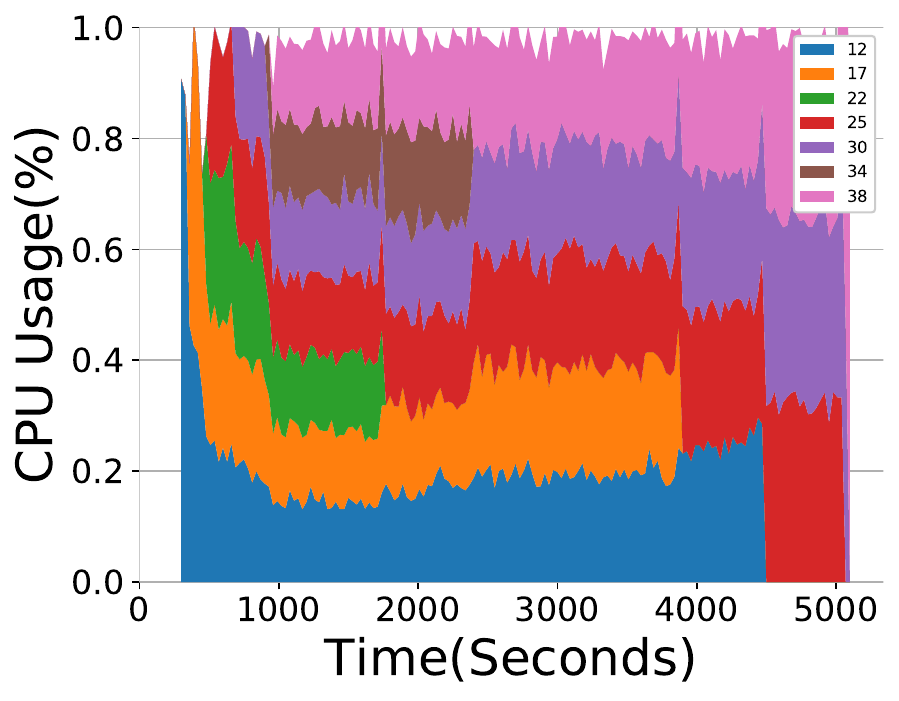}
\caption{Worker-1}
      \label{fig:docker:cluster:cpu1}
      \end{subfigure} 
      \begin{subfigure}[b]{0.24\textwidth}
\centering
         \includegraphics[width=\textwidth]{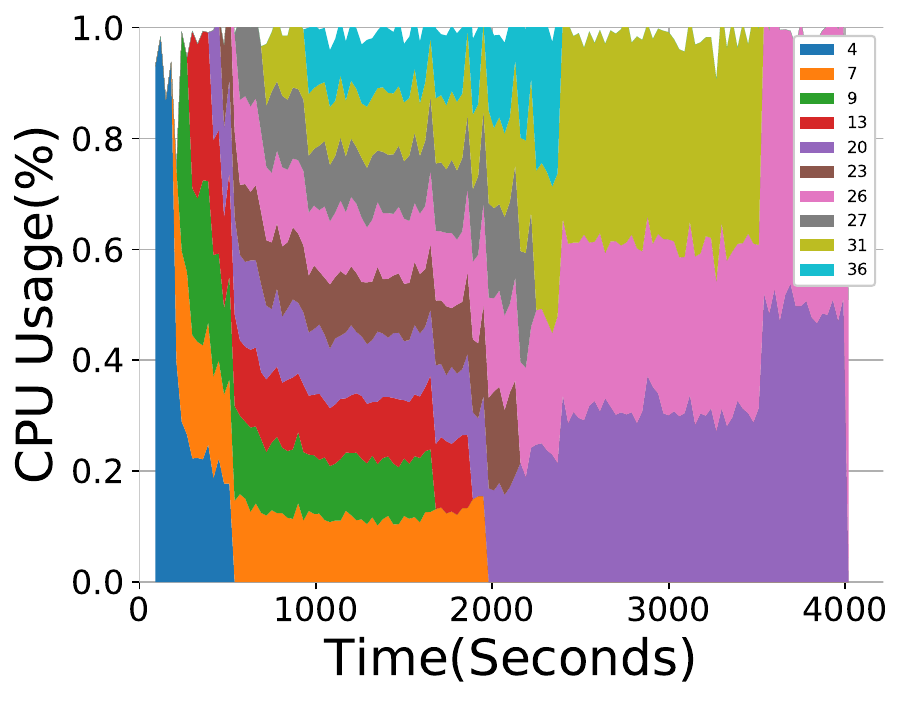}
\caption{Worker-2}
      \label{fig:docker:cluster:cpu2}
      \end{subfigure} %
      \begin{subfigure}[b]{0.24\textwidth}
\centering
         \includegraphics[width=\textwidth]{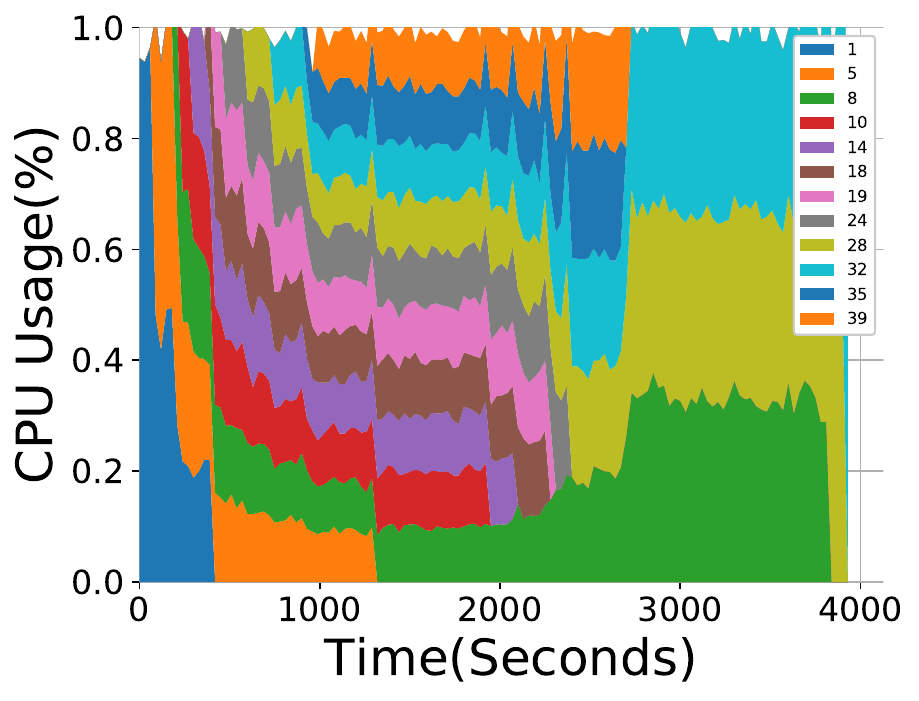}
\caption{Worker-3}
      \label{fig:docker:cluster:cpu3}
      \end{subfigure} %
      \begin{subfigure}[b]{0.24\textwidth}
	\centering
         \includegraphics[width=\textwidth]{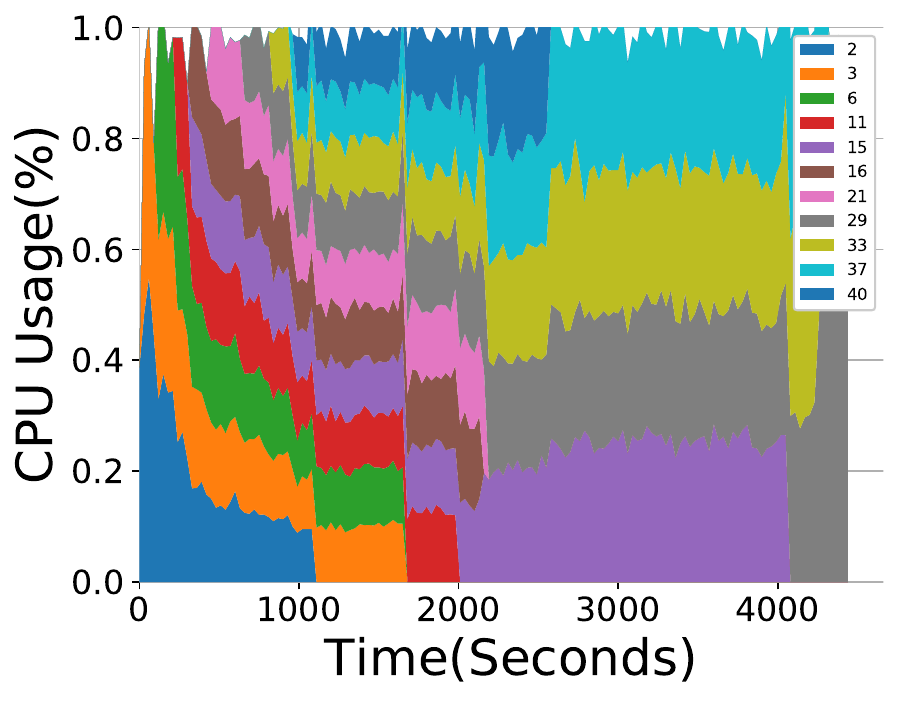}
	\caption{Worker-4}
      \label{fig:docker:cluster:cpu4}
      \end{subfigure} %
\caption{Container Placement with Docker Swarm}  
\label{cluster:docker:cpu}               
\end{figure*}

\begin{figure*}[h]
   \centering
         \begin{subfigure}[b]{0.24\textwidth}
\centering
         \includegraphics[width=\textwidth]{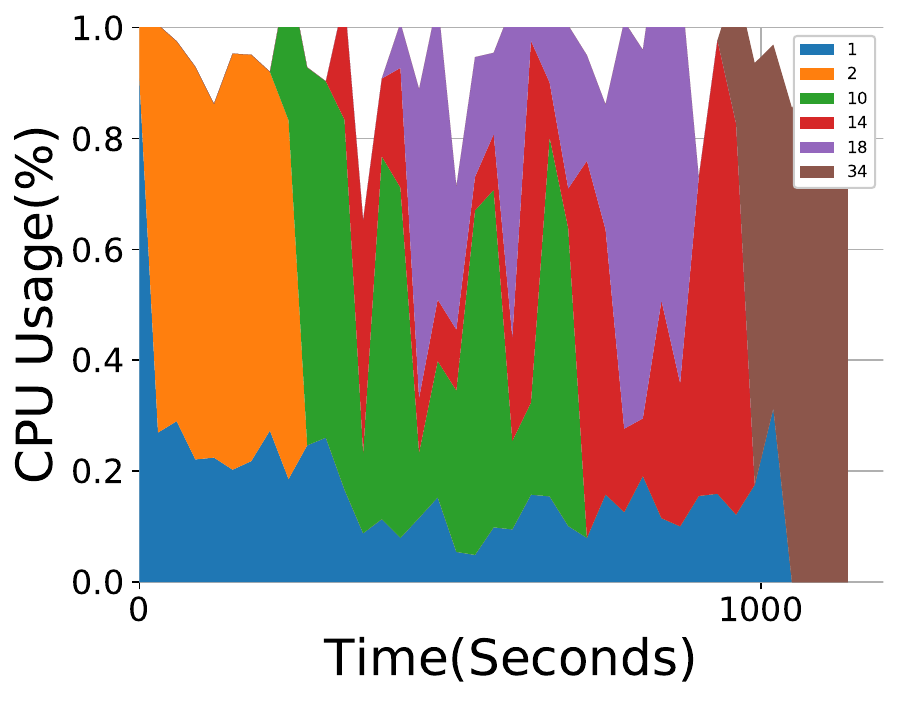}
\caption{Worker-1}
      \label{fig:k8s:cluster:cpu1}
      \end{subfigure} 
      \begin{subfigure}[b]{0.24\textwidth}
\centering
         \includegraphics[width=\textwidth]{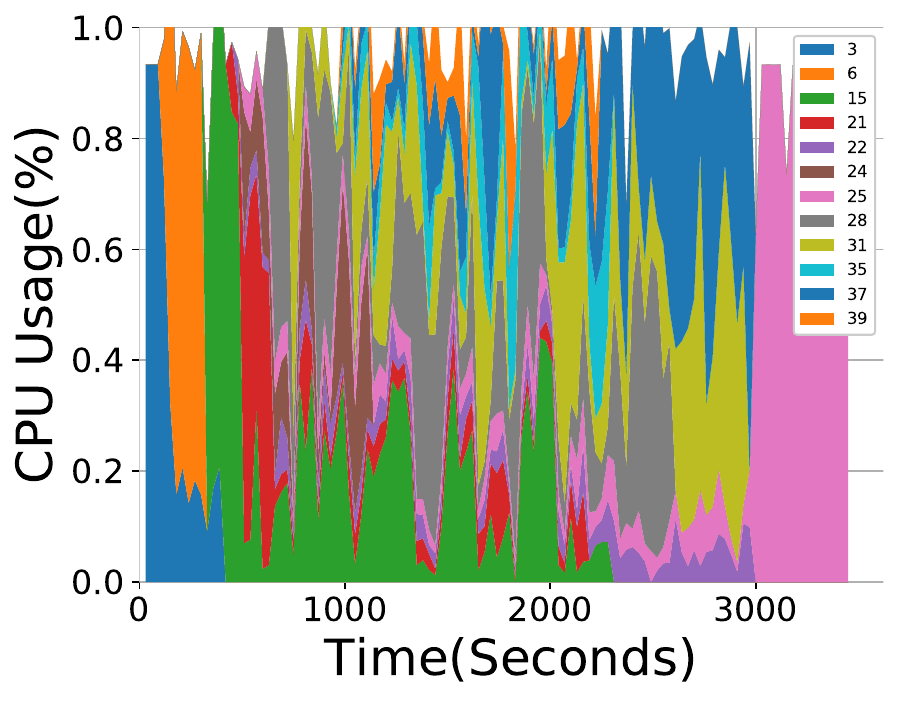}
\caption{Worker-2}
      \label{fig:k8s:cluster:cpu2}
      \end{subfigure} %
      \begin{subfigure}[b]{0.24\textwidth}
\centering
         \includegraphics[width=\textwidth]{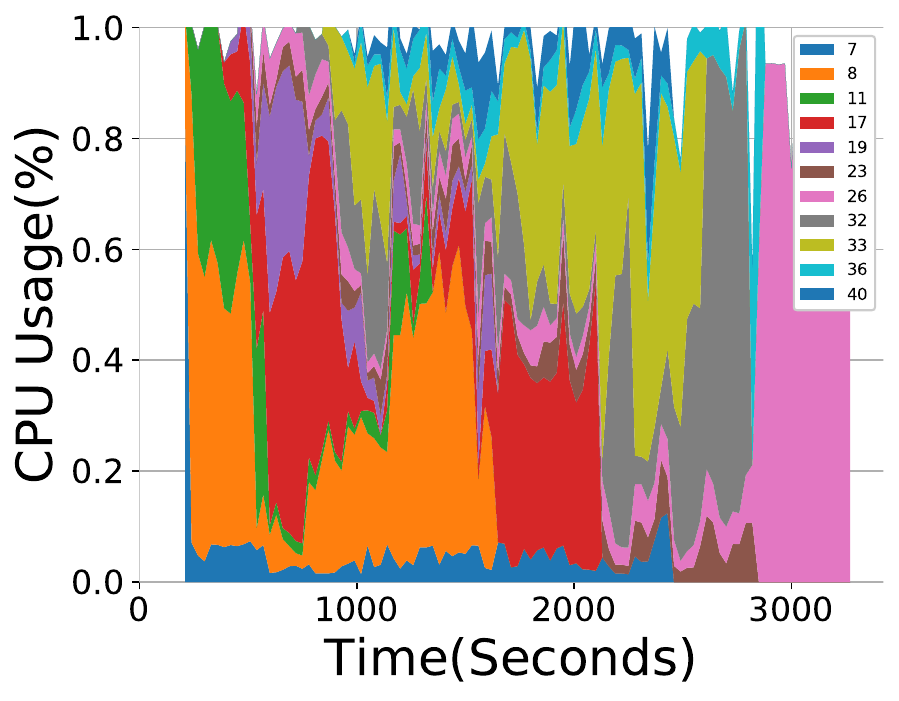}
\caption{Worker-3}
      \label{fig:k8s:cluster:cpu3}
      \end{subfigure} %
      \begin{subfigure}[b]{0.24\textwidth}
	\centering
         \includegraphics[width=\textwidth]{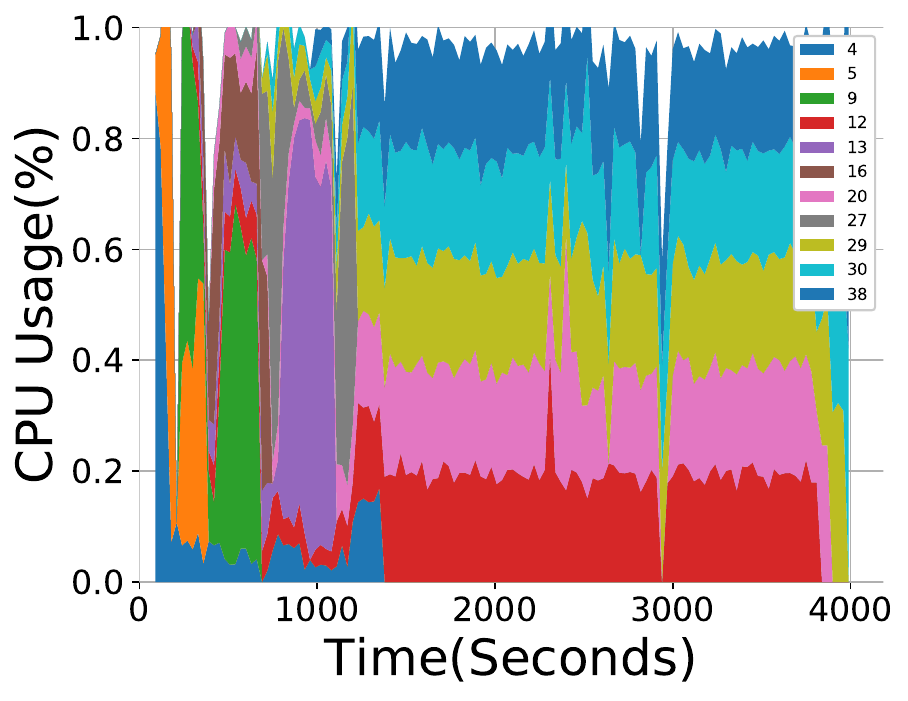}
	\caption{Worker-4}
      \label{fig:k8s:cluster:cpu4}
      \end{subfigure} %
\caption{Container Placement with Kubernetes}  
\label{cluster:k8s:cpu}               
\end{figure*}

\begin{figure*}[h]
   \centering
         \begin{subfigure}[b]{0.24\textwidth}
\centering
         \includegraphics[width=\textwidth]{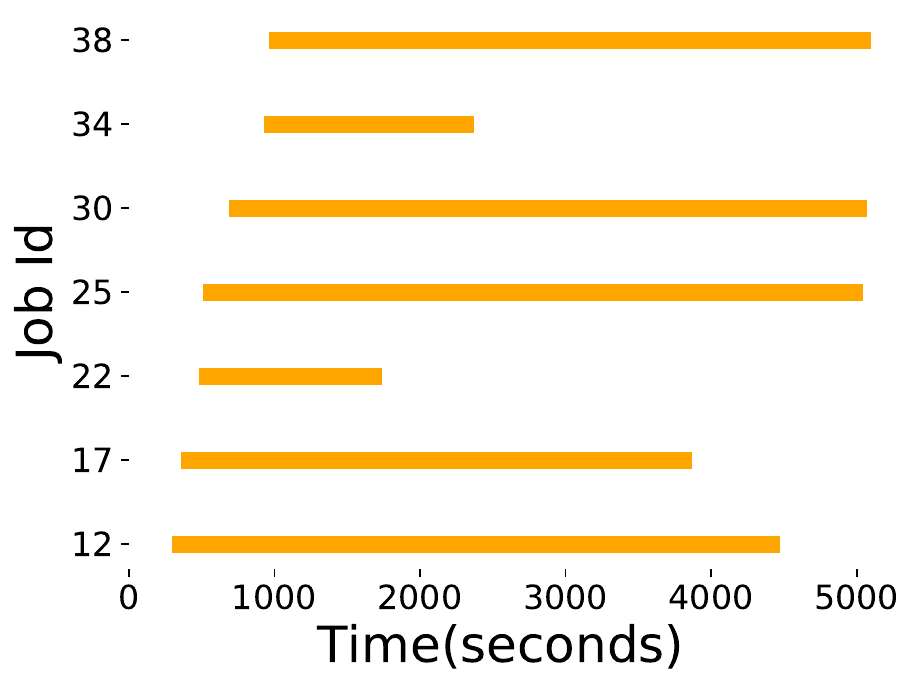}
\caption{Worker-1}
      \label{fig:docker:cluster:w1}
      \end{subfigure} 
      \begin{subfigure}[b]{0.24\textwidth}
\centering
         \includegraphics[width=\textwidth]{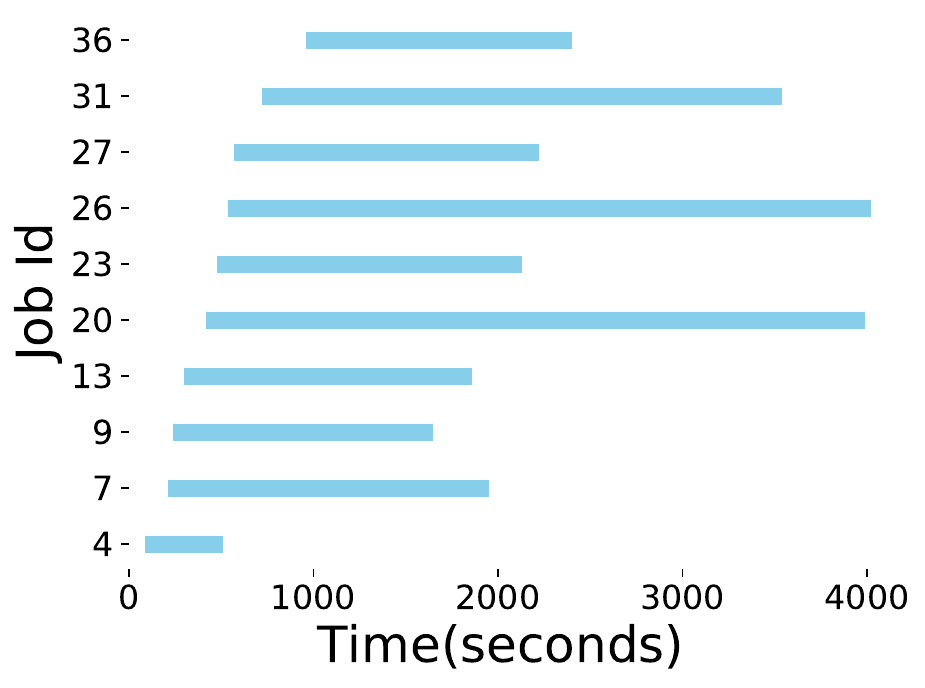}
\caption{Worker-2}
      \label{fig:docker:cluster:w2}
      \end{subfigure} %
      \begin{subfigure}[b]{0.24\textwidth}
\centering
         \includegraphics[width=\textwidth]{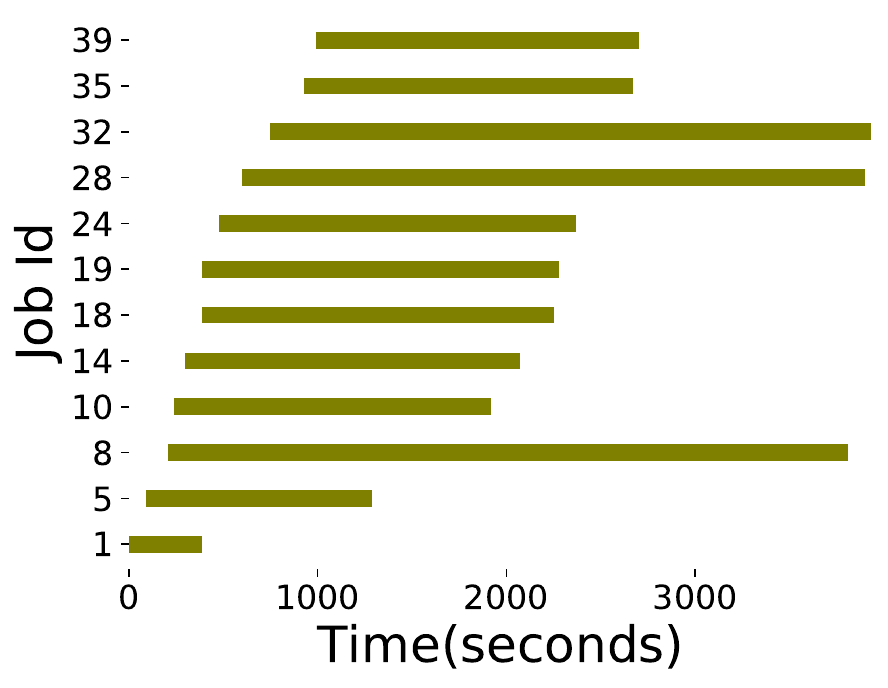}
\caption{Worker-3}
      \label{fig:docker:cluster:w3}
      \end{subfigure} %
      \begin{subfigure}[b]{0.24\textwidth}
	\centering
         \includegraphics[width=\textwidth]{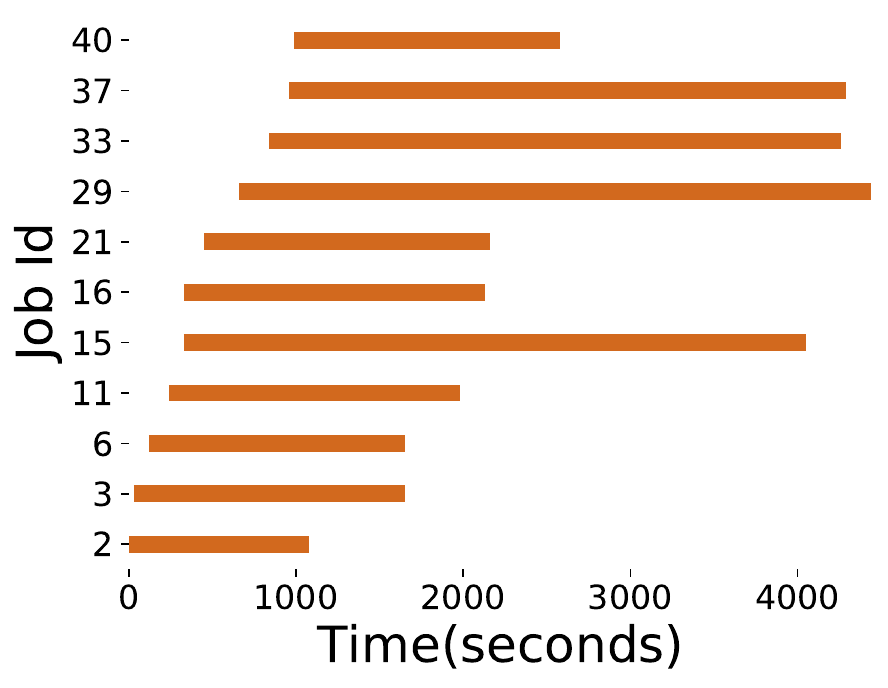}
	\caption{Worker-4}
      \label{fig:docker:cluster:w4}
      \end{subfigure} %
\caption{Container Placement with Docker Swarm}  
\label{cluster:docker}               
\end{figure*}

\begin{figure*}[!htb]
   \centering
         \begin{subfigure}[b]{0.24\textwidth}
\centering
         \includegraphics[width=\textwidth]{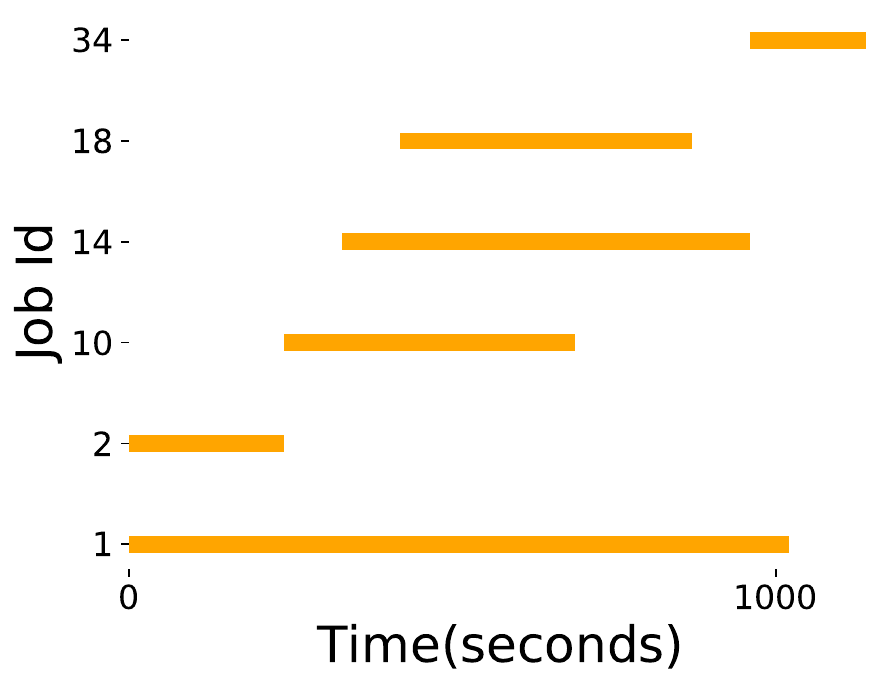}
\caption{Worker-1}
      \label{fig:k8s:cluster:w1}
      \end{subfigure} 
      \begin{subfigure}[b]{0.24\textwidth}
\centering
         \includegraphics[width=\textwidth]{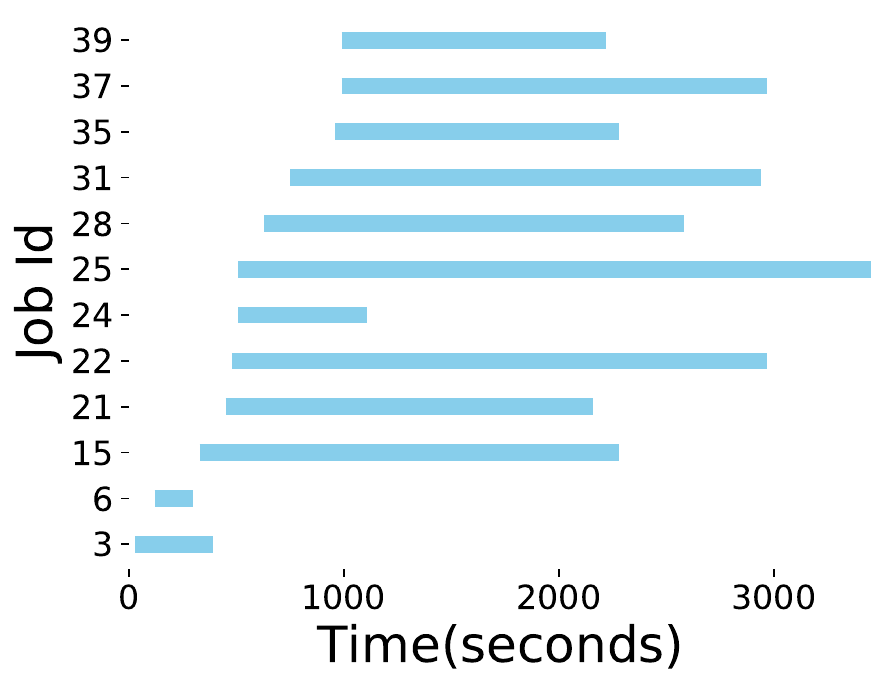}
\caption{Worker-2}
      \label{fig:k8s:cluster:w2}
      \end{subfigure} %
      \begin{subfigure}[b]{0.24\textwidth}
\centering
         \includegraphics[width=\textwidth]{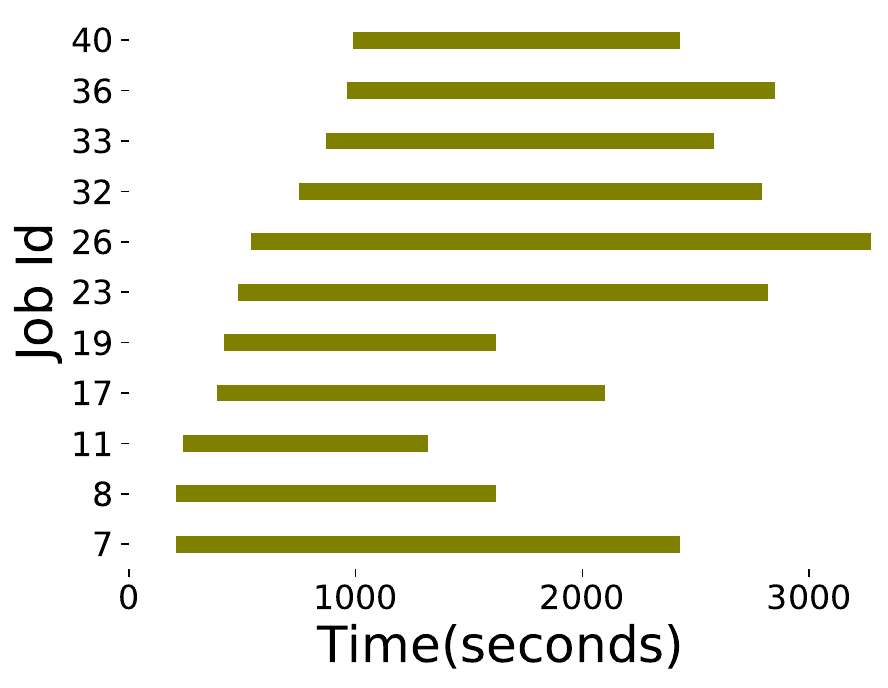}
\caption{Worker-3}
      \label{fig:k8s:cluster:w3}
      \end{subfigure} %
      \begin{subfigure}[b]{0.24\textwidth}
	\centering
         \includegraphics[width=\textwidth]{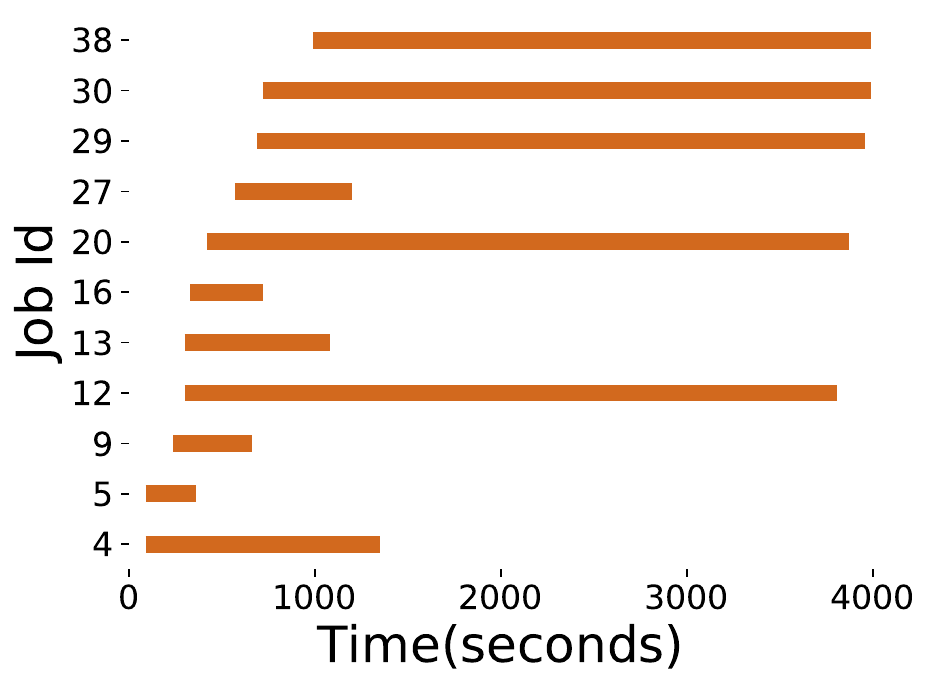}
	\caption{Worker-4}
      \label{fig:k8s:cluster:w4}
      \end{subfigure} %
\caption{Container Placement with Kubernetes}  
\label{cluster:k8s}               
\end{figure*}

Figure~\ref{cluster_bar} plots the results of completion time for each job in the cluster. Since deep learning applications are 
As shown in the figure, Kubernetes outperforms Docker Swarm in 34 out of  40 jobs. The average completion time is reduced from 
2328.5s to 1576.6s, 32.3\%. 
The largest gain is achieved on Job-8, which reduces 60.3\%, from 3605.7s to 1430.3s.  As expected, the improvement is due to dynamic resource adjustment to prevent idling tasks and increase resource utilization. Fig.~\ref{cluster:docker:cpu} and Fig.~\ref{cluster:k8s:cpu} prove the conclusion since a similar trend is found for both Docker Swarm and Kubernetes as we do in the single node system.

However, Docker Swarm wins significantly on 6 out of 40 jobs.  For example, the completion time of Job-22 is 2532.2s for Kubernetes
and 1287.1s with Docker Swarm, which is a 96.7\% increase.
This is due to the fact that, in a cluster environment, the system scheduler not only manages resources on each individual workers,
but also decides the workload on a specific worker through initial container placement.
With a deeper investigation of the system, we find that the container distribution on Docker Swarm and Kubernetes is different.
Fig.~\ref{cluster:docker} and Fig.~\ref{cluster:k8s} illustrate the container placement in this experiment. 
First of all, we discover that for both platforms, Worker-1 hosts less workload than other nodes in the cluster. This is because, in our setting, Worker-1 serves as both Manager that is responsible for system-wide scheduling tasks and Worker, which provides
the computing resources for the workloads. Therefore, different platforms have various algorithms to prevent overwhelming the manager. As we can see that Kubernetes distribute the first two jobs (submitted at 4s and 15s) to Worker-1 since, at the very beginning, the manager only involves very limited tasks and system trade it as a normal worker. When more tasks join the cluster,
the number of scheduling tasks that the manager handles increases. Therefore, the workload on the manager maintains at the same level until 248s, when Job-10 is submitted. With Docker Swarm, however, it tries to protect its manager from the very beginning. 
The first container that assigns to Worker-1 is Job-12 at 302s. Consequently, the Worker-1 is idling from 0 to 301s, which leads
to significant performance differences.

\section{Conclusion}
\label{con}
In this project, we explored the system performance of two cloud native platforms, Docker and Kubernetes.
Together with Prometheus and Grafana, we built a container monitor system to keep tracking the resource usage of each job
on the worker nodes.
We conducted intensive experiments based on a testbed hosted by the datacenter of NSF Cloudlab.
Focusing on two popular computing workloads, deep learning (Tensorflow and Pytorch) and big data (Apache Yarn and Spark) applications, our experiments utilize burst, fixed, and random submission schedules on both single node and cluster environment.
The results demonstrate that considering each individual platform, the completion time was reduced for up to 79.4\% and 69.4\% by changing the default configurations on Docker and Kubernetes. Comparing two platforms, 
we discover that
Docker platform delayed resource release process for up
to 116.7\% for short-lived deep learning jobs. Furthermore, the completion time increased up to 96.7\%
in the cluster environment due to different resource management schemes.

As future work, we will consider designing algorithms to dynamically adjust the thread numbers for Tensorflow and Pytorch
to boost system performance by increase resource utilization. In addition, we plan to explore the system performance 
from the perspective of container recovery and migration to unfold an optimized container management scheme. Finally, 
we intend to conduct experiments to understand the system performance of the two platforms with long-lived workloads, e.g. web and database services.


\bibliographystyle{IEEEtran}

\begin{thebibliography}{10}
\providecommand{\url}[1]{#1}
\csname url@samestyle\endcsname
\providecommand{\newblock}{\relax}
\providecommand{\bibinfo}[2]{#2}
\providecommand{\BIBentrySTDinterwordspacing}{\spaceskip=0pt\relax}
\providecommand{\BIBentryALTinterwordstretchfactor}{4}
\providecommand{\BIBentryALTinterwordspacing}{\spaceskip=\fontdimen2\font plus
\BIBentryALTinterwordstretchfactor\fontdimen3\font minus
  \fontdimen4\font\relax}
\providecommand{\BIBforeignlanguage}[2]{{%
\expandafter\ifx\csname l@#1\endcsname\relax
\typeout{** WARNING: IEEEtran.bst: No hyphenation pattern has been}%
\typeout{** loaded for the language `#1'. Using the pattern for}%
\typeout{** the default language instead.}%
\else
\language=\csname l@#1\endcsname
\fi
#2}}
\providecommand{\BIBdecl}{\relax}
\BIBdecl

\bibitem{forbes}
``Forbes report,''
  {\url{https://www.forbes.com/sites/louiscolumbus/2015/01/24/roundup-of-cloud-computing-forecasts-and-market-estimates-2015}}.

\bibitem{aws}
``Amazon web service,'' {\url{https://aws.amazon.com/}}.

\bibitem{azure}
``Microsoft azure,'' {\url{https://azure.microsoft.com/en-us/}}.

\bibitem{gcp}
``Google cloud platform,'' {\url{https://cloud.google.com/}}.

\bibitem{docker}
``Docker,'' {\url{https://www.docker.com/}}.

\bibitem{dockerswarm}
``Docker swarm,'' {\url{https://docs.docker.com/engine/swarm/}}.

\bibitem{k8s}
``kubernetes,'' {\url{https://kubernetes.io/}}.

\bibitem{awshistory}
``Aws history,'' {\url{https://mediatemple.net/blog/news/brief-history-aws/}}.

\bibitem{fb}
``Facebook,'' {\url{https://facebook.com/}}.

\bibitem{twitter}
``Twitter,'' {\url{https://twitter.com/}}.

\bibitem{gmail}
``Google gmail,'' {\url{https://gmail.com}}.

\bibitem{outlook}
``Outlook,'' {\url{https://outlook.com/}}.

\bibitem{chen2020woa}
X.~Chen, L.~Cheng, C.~Liu, Q.~Liu, J.~Liu, Y.~Mao, and J.~Murphy, ``A woa-based
  optimization approach for task scheduling in cloud computing systems,''
  \emph{IEEE Systems Journal}, 2020.

\bibitem{mao2017draps}
Y.~Mao, J.~Oak, A.~Pompili, D.~Beer, T.~Han, and P.~Hu, ``Draps: Dynamic and
  resource-aware placement scheme for docker containers in a heterogeneous
  cluster,'' in \emph{2017 IEEE 36th International Performance Computing and
  Communications Conference (IPCCC)}.\hskip 1em plus 0.5em minus 0.4em\relax
  IEEE, 2017, pp. 1--8.

\bibitem{mao2014pasa}
Y.~Mao, J.~Wang, J.~P. Cohen, and B.~Sheng, ``Pasa: Passive broadcast for
  smartphone ad-hoc networks,'' in \emph{2014 23rd International Conference on
  Computer Communication and Networks (ICCCN)}.\hskip 1em plus 0.5em minus
  0.4em\relax IEEE, 2014, pp. 1--8.

\bibitem{mao2018dress}
Y.~Mao, V.~Green, J.~Wang, H.~Xiong, and Z.~Guo, ``Dress: Dynamic
  resource-reservation scheme for congested data-intensive computing
  platforms,'' in \emph{2018 IEEE 11th International Conference on Cloud
  Computing (CLOUD)}.\hskip 1em plus 0.5em minus 0.4em\relax IEEE, 2018, pp.
  694--701.

\bibitem{acharya2019workload}
A.~Acharya, Y.~Hou, Y.~Mao, M.~Xian, and J.~Yuan, ``Workload-aware task
  placement in edge-assisted human re-identification,'' in \emph{2019 16th
  Annual IEEE International Conference on Sensing, Communication, and
  Networking (SECON)}.\hskip 1em plus 0.5em minus 0.4em\relax IEEE, 2019, pp.
  1--9.

\bibitem{acharya2019edge}
A.~Acharya, Y.~Hou, Y.~Mao, and J.~Yuan, ``Edge-assisted image processing with
  joint optimization of responding and placement strategy,'' in \emph{2019
  International Conference on Internet of Things (iThings) and IEEE Green
  Computing and Communications (GreenCom) and IEEE Cyber, Physical and Social
  Computing (CPSCom) and IEEE Smart Data (SmartData)}.\hskip 1em plus 0.5em
  minus 0.4em\relax IEEE, 2019, pp. 1241--1248.

\bibitem{harvey2017edos}
H.~H. Harvey, Y.~Mao, Y.~Hou, and B.~Sheng, ``Edos: Edge assisted offloading
  system for mobile devices,'' in \emph{2017 26th International Conference on
  Computer Communication and Networks (ICCCN)}.\hskip 1em plus 0.5em minus
  0.4em\relax IEEE, 2017, pp. 1--9.

\bibitem{xu2013managing}
F.~Xu, F.~Liu, H.~Jin, and A.~V. Vasilakos, ``Managing performance overhead of
  virtual machines in cloud computing: A survey, state of the art, and future
  directions,'' \emph{Proceedings of the IEEE}, vol. 102, no.~1, pp. 11--31,
  2013.

\bibitem{felter2015updated}
W.~Felter, A.~Ferreira, R.~Rajamony, and J.~Rubio, ``An updated performance
  comparison of virtual machines and linux containers,'' in \emph{2015 IEEE
  international symposium on performance analysis of systems and software
  (ISPASS)}.\hskip 1em plus 0.5em minus 0.4em\relax IEEE, 2015, pp. 171--172.

\bibitem{sharma2016containers}
P.~Sharma, L.~Chaufournier, P.~Shenoy, and Y.~Tay, ``Containers and virtual
  machines at scale: A comparative study,'' in \emph{Proceedings of the 17th
  International Middleware Conference}.\hskip 1em plus 0.5em minus 0.4em\relax
  ACM, 2016, p.~1.

\bibitem{piraghaj2017containercloudsim}
S.~F. Piraghaj, A.~V. Dastjerdi, R.~N. Calheiros, and R.~Buyya,
  ``Containercloudsim: An environment for modeling and simulation of containers
  in cloud data centers,'' \emph{Software: Practice and Experience}, vol.~47,
  no.~4, pp. 505--521, 2017.

\bibitem{tay2017performance}
Y.~Tay, K.~Gaurav, and P.~Karkun, ``A performance comparison of containers and
  virtual machines in workload migration context,'' in \emph{2017 IEEE 37th
  International Conference on Distributed Computing Systems Workshops
  (ICDCSW)}.\hskip 1em plus 0.5em minus 0.4em\relax IEEE, 2017, pp. 61--66.

\bibitem{bhimani2017accelerating}
J.~Bhimani, Z.~Yang, M.~Leeser, and N.~Mi, ``Accelerating big data applications
  using lightweight virtualization framework on enterprise cloud,'' in
  \emph{2017 IEEE High Performance Extreme Computing Conference (HPEC)}.\hskip
  1em plus 0.5em minus 0.4em\relax IEEE, 2017, pp. 1--7.

\bibitem{zhang2018comparative}
Q.~Zhang, L.~Liu, C.~Pu, Q.~Dou, L.~Wu, and W.~Zhou, ``A comparative study of
  containers and virtual machines in big data environment,'' \emph{arXiv
  preprint arXiv:1807.01842}, 2018.

\bibitem{khan2017key}
A.~Khan, ``Key characteristics of a container orchestration platform to enable
  a modern application,'' \emph{IEEE Cloud Computing}, vol.~4, no.~5, pp.
  42--48, 2017.

\bibitem{xu2018nbwguard}
C.~Xu, K.~Rajamani, and W.~Felter, ``Nbwguard: Realizing network qos for
  kubernetes,'' in \emph{Proceedings of the 19th International Middleware
  Conference Industry}.\hskip 1em plus 0.5em minus 0.4em\relax ACM, 2018, pp.
  32--38.

\bibitem{fu2019progress}
Y.~Fu, S.~Zhang, J.~Terrero, Y.~Mao, G.~Liu, S.~Li, and D.~Tao,
  ``Progress-based container scheduling for short-lived applications in a
  kubernetes cluster,'' in \emph{2019 IEEE International Conference on Big Data
  (Big Data)}.\hskip 1em plus 0.5em minus 0.4em\relax IEEE, 2019, pp. 278--287.

\bibitem{zheng2019target}
W.~Zheng, Y.~Song, Z.~Guo, Y.~Cui, S.~Gu, Y.~Mao, and L.~Cheng, ``Target-based
  resource allocation for deep learning applications in a multi-tenancy
  system,'' in \emph{2019 IEEE High Performance Extreme Computing Conference
  (HPEC)}.\hskip 1em plus 0.5em minus 0.4em\relax IEEE, 2019, pp. 1--7.

\bibitem{zheng2019flowcon}
W.~Zheng, M.~Tynes, H.~Gorelick, Y.~Mao, L.~Cheng, and Y.~Hou, ``Flowcon:
  Elastic flow configuration for containerized deep learning applications,'' in
  \emph{Proceedings of the 48th International Conference on Parallel
  Processing}, 2019, pp. 1--10.

\bibitem{tensorflow}
``Tensorflow,'' {\url{https://www.tensorflow.org/}}.

\bibitem{pytorch}
``Pytorch,'' {\url{https://pytorch.org/}}.

\bibitem{spark}
``Apache spark,'' {\url{https://spark.apache.org/}}.

\bibitem{cncf}
``Cloud native computing foundation,'' {\url{https://www.cncf.io/}}.

\bibitem{dockerhub}
``Docker hub,'' {\url{https://hub.docker.com/}}.

\bibitem{yaml}
``Yaml,'' {\url{https://yaml.org/}}.

\bibitem{prometheus}
``Prometheus,'' {\url{https://prometheus.io}}.

\bibitem{grafana}
``Grafana,'' {\url{https://grafana.com}}.

\bibitem{yarn}
``Hadoop yarn,''
  {\url{https://hadoop.apache.org/docs/current/hadoop-yarn/hadoop-yarn-site/YARN.html}}.

\bibitem{cloudlab}
``Nsf cloudlab,'' {\url{https://cloudlab.us/}}.

\end{thebibliography}

\end{document}